\documentclass[journal]{IEEEtran}
\usepackage{float}
\usepackage{subfigure}
\usepackage{graphicx}
\usepackage[linkcolor=red]{hyperref}
\usepackage[cmex10]{amsmath}
\usepackage{amssymb}
\usepackage{bm}
\usepackage{array}
\usepackage[numbers,sort&compress]{natbib} %reference
\usepackage{makecell}
\usepackage{multirow}
\usepackage{amsthm,amsmath,amssymb}
\usepackage{bbding} % 添加对号叉号
\usepackage{mathrsfs}
\usepackage{enumerate}
\usepackage{amsmath,epsfig}
\usepackage[table]{xcolor}
\usepackage{booktabs}
\usepackage{color}
\usepackage{xcolor}
\usepackage{url}
\usepackage{diagbox}
\usepackage{ragged2e}

\usepackage[ruled,linesnumbered]{algorithm2e}

\begin{document}
\title{A Robust Image Steganographic Scheme against General Scaling Attacks}
%\author{Qingliang~Liu, Jiangqun~Ni*,~\IEEEmembership{Member,~IEEE}, Weizhe Zhang, ~\IEEEmembership{Senior Member,~IEEE}, Xiangyang Luo, and Jiwu Huang, ~\IEEEmembership{Fellow,~IEEE} % <-this % stops a space
\author{Qingliang~Liu, Jiangqun~Ni*, Weizhe Zhang, Xiangyang Luo, and Jiwu Huang % <-this % stops a
%\thanks{This work was supported by xxxxx}
%\thanks{Q. Liu, M. Jian and X. Hu are with the School of Computer Science and Engineering, Sun Yat-sen University, Guangzhou~510006, China~(e-mail:~liuqliang3@mail2.sysu.edu.cn;~jianmx@mail2.sysu.edu.cn; {huxlei3@mail.sysu.edu.cn}).}% <-this % stops a space
%\thanks{J. Ni is with the School of Computer Science and Engineering, Sun Yat-sen University, Guangzhou~510006, China, and is also with Cyberspace Security Research Center, Peng Cheng Laboratory, Shenzhen~518055, China~(e-mail:~issjqni@mail.sysu.edu.cn).}% <-this % stops a space
%\thanks{* Corresponding author}
}

\maketitle

%As a general rule, do not put math, special symbols or citations
\begin{abstract}
Conventional covert image communication is assumed to transmit the message, in the securest way possible for a given payload, over lossless channels, and the associated steganographic schemes are generally vulnerable to active attacks, e.g., JPEG re-compression, scaling, as seen on social networks. 
Although considerable progress has been made on robust steganography against JPEG re-compression, there exist few steganographic schemes capable of resisting to scaling attacks due to the tricky inverse interpolations involved in algorithm design.  
To tackle this issue, a framework for robust image steganography resisting to scaling with general interpolations either in standard form with fixed interpolation block, or pre-filtering based anti-aliasing implementation with variable block, is proposed in this paper.
And the task of robust steganography can be formulated as the one of constrained integer programming aiming at perfectly recover the secret message from stego image while minimizing the difference between cover and stego images.
By introducing a metric - the degree of pixel involvement (dPI) to identify the modifiable pixels in cover image, the optimization problem above could be effectively solved using branch and bound algorithm (B\&B). 
In addition, a customized distortion function for scaled stego images is adopted to further boost the security performance. 
Extensive experiments are carried out which demonstrate that the proposed scheme could not only outperform the prior art in terms of security by a clear margin, but also be applicable to resisting the scaling attacks with various interpolation techniques at arbitrary scaling factors (SFs).

\end{abstract}


\IEEEpeerreviewmaketitle
\section{Introduction}

\IEEEPARstart {I}{mage} steganography is the science and art of covert communication, in which the sender embeds the secret message into the cover images by slightly modifying the pixel values (in spatial domain) or the quantized DCT coefficients (in JPEG domain). 
To conceal the very existence of communication, the stego images have to be statistically indistinguishable from the cover images.

Traditionally, image steganography assumes a lossless channel without active attacks, e.g., JPEG re-compression, scaling, or the combination of the two. And the steganographic schemes associated with the image covert communication should have acceptable statistical imperceptibility (security) with sufficient payload, although these two objectives are generally conflicting with each other. 
The modern adaptive image steganographic schemes under the framework of minimal distortion embedding \cite{filler2011minimizing} work well for lossless channels, they, however, could by no means survive the active attacks experienced in some covert communications over the lossy channel, e.g., the applications on social networks (Facebook, Twitter, Wechat etc.), where the uploaded images are generally resized and JPEG re-compressed.  
Therefore, there is an urgent need to develop effective image steganographic schemes that are capable of resisting to both statistical detection and known active attacks at the same time.

The security performance in terms of statistical undetectability has long been considered as one of the key design objectives for stego systems over the lossless channel, which is generally evaluated in the form of decision error of steganalyzers trained on stego and cover images.
The past decade has witnessed the emergence of a large number of content-adaptive image steganographic methods, either heuristically designed or model-driven.  
According to the domain in which the secret messages are embedded, these methods can be categorized as the ones in spatial domain and JPEG domain, respectively.  
For image steganography in spatial domain, the messages are embedded into the complex regions with rich texture contents in cover images, such as WOW \cite{holub2012designing}, S-UNIWARD \cite{holub2014universal}, HiLL \cite{li2014new}, MiPOD \cite{Sedighi2016content}, GMRF \cite{su2021image} etc. 
While for JPEG steganography, the quantized DCT coefficients corresponding to complex texture regions, which are more tolerant to steganalytic attacks, are utilized in priority for data embedding, such as UED \cite{guo2014uniform}, UERD \cite{guo2015using}, J-UNIWARD \cite{holub2014universal}, BET \cite{hu2018efficient}, J-MiPOD \cite{R2022efficient} etc.

JPEG re-compression, scaling and their combination are the most common active attacks on lossy steganographic channels, as often found in social networks. 
For JPEG re-compression, there are plenty of robust steganographic schemes against known JPEG compression, which aims not only to ensure the statistical indistinguishability of the stego images from the cover images (security),  but also to perfectly recover the embedded messages from the stegos (robustness).  
And this is achieved by transport channel matching \cite{zhao2019improving}, by fitting the inverse procedure of the JPEG compression channel with a pre-trained auto-encoder \cite{lu2021secure} etc. 
For scaling attacks, however, there exist few steganographic schemes capable of resisting to scaling attacks due to the tricky inverse interpolations involved in the algorithm design. 
Generally speaking, image scaling operation (either up-scaling or down-scaling) involves image resampling with interpolation. 
Unless otherwise specified, only the down-scaling attacks with scaling factor (SF) in $(0,1]$ are addressed throughout the paper, as they are the ones the social networks run into most frequently.
In general, image scaling or resize operations involve standard interpolation methods with fixed block sizes, such as nearest, bilinear (the closest $2 \times 2$ neighborhood) and bicubic (the closest $4 \times 4$ neighborhood), and their anti-aliasing variants of latter two with variable block sizes, which are the subroutines invoked from  MATLAB \cite{matlabresize} or OpenCV \cite{opencvreize}. 
Early anti-scaling steganographic schemes \cite{zhang2018an, zhang2019zernike} heavily depend on watermarking techniques, albeit being capable of surviving the scaling attacks, they usually exhibit low embedding capacity and negligible security performance.  
Recently, Zhu et al. \cite{zhang2018image} proposed a robust image steganographic scheme against nearest neighbor interpolation scaling attack by taking advantage of the invariant pixels to embed messages. 
Later, they further extend their method to the more general scaling attacks with anti-aliasing bilinear and bicubic interpolations by incorporating the inverse interpolation rule from scaled images to original cover images.
Although relatively large embedding capacity and satisfactory statistical undetectability are achieved for scaled stego images, the improvement of the security performance for the generated stego images at the same dimension as the original images (the proxy stego images in this paper) is not evident due to the selection of modifiable pixels in cover images in the inverse interpolation process, which will be discussed later in Section \ref{Related}. 
For the same reason, the method in \cite{zhu2021inverse} could not be used for standard bilinear and bicubic scaling attacks and the SFs for anti-aliasing bilinear and bicubic interpolation scaling are confined to (0,0.5] and (0,0.25], respectively. 

\begin{figure*}[t]
\centering
\centerline{\epsfig{figure=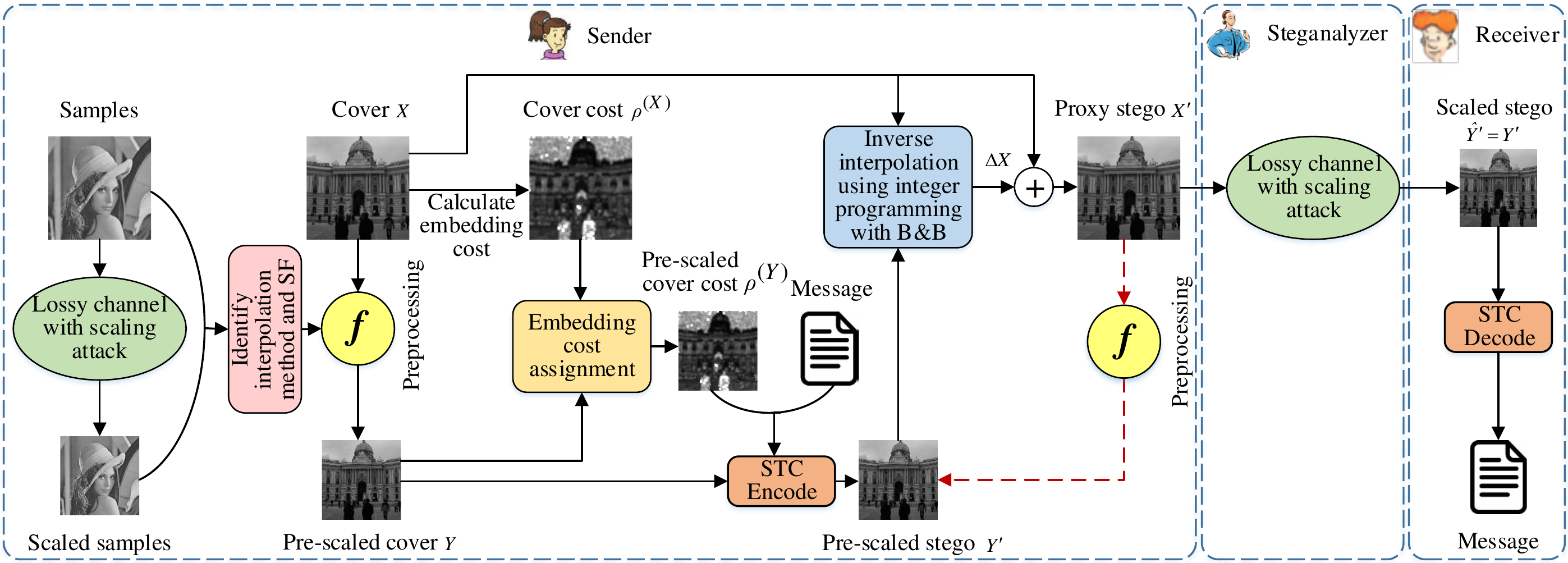,width=1.0\linewidth}}
\caption{The diagram of the proposed robust steganographic framework against general scaling attacks.} 
\label{Fig:1}
\end{figure*}

In this paper, we propose a flexible framework for robust image steganography against general scaling attacks at arbitrary scaling factors (SFs). 
For steganographic image covert communication over the lossy channel with known general scaling attacks as shown in Fig. \ref{Fig:1}, the cover image $X$ is resized to the scaled cover image $Y$, in which the message is embedded to generate the scaled stego image $Y^{\prime}$. 
The proxy stego image with the same dimension as cover image $X$, i.e.,  $X^{\prime} = X + \Delta\!X$ is sought based on $Y^{\prime}$ through inverse interpolation with the objective to minimize
$\Vert X - X^{\prime} \Vert_1$ for the possibly best security performance and reliably recover the message from $f(x^{\prime})$ (robustness), where $f$ is the function of involved interpolation scaling.   
For each embedding variation $\pm 1$ in scaled stego image $Y^{\prime}$, a constraint is established between the marked pixel and its corresponding pixels involved in the interpolation computation in proxy stego image $X^{\prime}$. 
If a pixel in proxy stego image is related to multiple constraints in inverse interpolation, an overdetermined constraint equation set is built, which is usually hard to solve or even unsolvable.
Another even bigger issue of concern in inverse interpolation is that, if each pixel in scaled cover image is determined as the embeddable pixel, the pixels located in the boundary of interpolation block of the proxy stego image are usually enforced to be modified to meet the embedding change in scaled stego image. 
Note that the weights in interpolation block, especially for the anti-aliasing implementation, are usually bell-shaped distribution, i.e., the weights in the interpolation block decrease as moving away from the center, thus leading to larger variations of the associated pixels in proxy stego image in response to the embedding changes, which is undesirable from the perspective of security. 
In our work, to tackle these issues,  the embeddable pixels in scaled cover images are elaborately determined, and a metric known as the degree of pixel involvement (dPI) is introduced to identify the modifiable pixels $\Delta x_{i,j} \neq 0$ in a cover image, so that: 
(1) the established constraint equation sets for inverse interpolation could be effectively solved; and 
(2) the modifications in the proxy stego image corresponding to the embedding changes take place in the central area of the interpolation block to keep the variations as small as possible compared to the cover image. 
Based on these, the task of robust steganography against scaling attacks in terms of inverse interpolation is then formulated as the one of constrained integer programming aiming at perfectly recovering the secret message from stego image while minimizing the difference between cover and proxy stego images, which could be effectively solved by exploring the Branch and Bound algorithm (B\&B). 
Finally, instead of determining the embedding costs directly based on the scaled cover image $Y$ in previous approaches, the distortion cost function in our work is derived from the cover image $X$ itself to take full advantage of the statistics of the cover image, which is shown to be able to further boost the security performance of proxy stego image $X^{\prime}$.

The effectiveness of the proposed robust image steganographic scheme against known scaling attacks is verified with evidence from extensive experiments using both standard bilinear/bicubic interpolation scaling and their anti-aliasing variants at arbitrary SFs, and for several state-of-the-art steganalyzers with a wide variety of payloads. 
The proposed scheme is shown not only to be capable of surviving the current predominant scaling channels with sufficiently large payloads, but also outperforms the prior arts in terms of security performance by a clear margin.

The remainder of this paper is organized as follows. 
In Section \ref{Related}, the key notations throughout the paper are defined, and the interpolation scaling method involved in the paper and the related works are briefly reviewed. The proposed robust image steganographic framework against general scaling attacks is described in Section \ref{ProposedMethod}, which is followed by the experimental results and analysis in Section IV. Finally, the concluding remarks are drawn in Section V.  

\section{Preliminaries and Related Works} \label{Related}
\subsection{Notations and Conventions}
Throughout the paper, a scalar is written in italicized lowercase, and the scalar matrix is represented in italicized uppercase letter unless otherwise specified.
For convenience, the major symbols adopted in this paper are summarized in Table \ref{tab:notation}.

\begin{table}[h]
\centering
\caption{Major notations.}
\label{tab:notation}
\begin{tabular}{cp{5.5cm}cp{3cm}}
\toprule[1 pt]
Notation & Description\\
\midrule

$X = [x_{i,j}]$ & The cover image.\\
$\rho^{(X)} = [\rho_{i,j}^{(x)}]$ & The distortion cost for cover image.\\
$Y = [y_{u,v}]$ & The pre-scaled cover image.\\
$\rho^{(Y)} = [\rho_{u,v}^{(y)}]$ & The distortion cost for pre-scaled cover image.\\
$Y^{\prime} = [y^{\prime}_{u,vv}]$ & The pre-scaled stego image.\\
$X^{\prime} = [x^{\prime}_{i,j}]$ & The proxy stego image.\\
$\hat{Y}^{\prime} = [\hat{y}^{\prime}_{u,v}]$ & The scaled stego image.\\
$f$ & The function of involved scaling channel.\\
$S\!F$ & The scaling factor for the scaling channel.\\
\multirow{2}{*}{$\boldsymbol{\mathbb{N}}_{k}$} & A set of integers, $\mathbb{N}_{k} = \{0,1,\cdots,k-1\}$, \newline where $k$ is a positive integer.\\
\bottomrule[1 pt]
\end{tabular}
\end{table}

\subsection{Image Interpolation Scaling}
For image steganographic covert communication on social network applications, e.g., Facebook, Twitter etc., the involved scaling attacks are usually down-scaling with interpolations. Therefore, in this paper, we only concentrate on the down-scaling channels with the SFs confined to $(0,1]$, unless otherwise specified.

\begin{figure}[t]
\centering
\centerline{\epsfig{figure=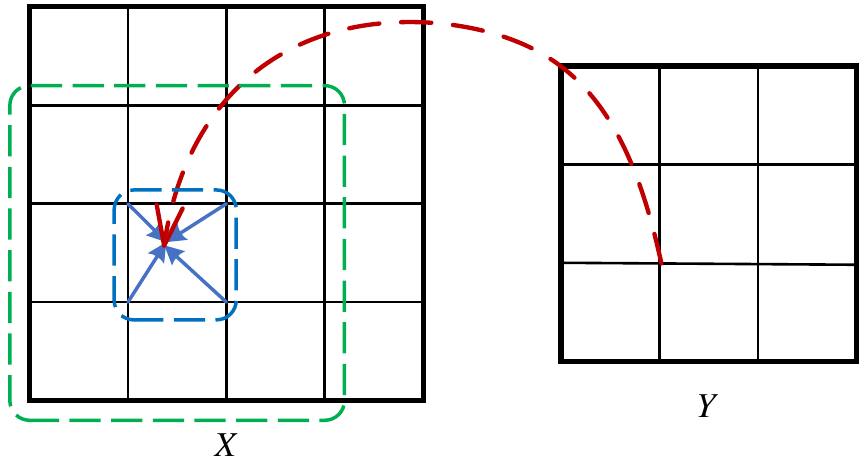,width=1\linewidth}}
\caption{The schematic diagram of image scaling with backward mapping and bilinear or bicubic interpolations, where the blue box of $2 \times 2$ and the larger green box correspond to the supporting box and interpolation box, respectively.} 
\label{Fig:2}
\end{figure}

Image scaling operation consists of geometric transformation and interpolations, e.g., nearest, bilinear, and bicubic either in standard forms with fixed block sizes, or their anti-aliasing variants with variable block sizes.
Note that anti-aliasing attempts to minimize the appearance of jagged diagonal edges. 
It works by taking into account of how much an ideal edge overlaps adjacent pixels and is widely adopted in social networks. 
For a given image $X = [x_{i,j}]$, it is shrunk with interpolation to generate the scaled image $Y = [y_{u,v}]$.
In general, backward mapping is adopted to re-sample $y_{u,v}$ in $Y$ based on the closest neighbors $P = [x_{i,j}]$  (interpolation block) in $X$ to $T^{-1}(u,v)$, $T^{-1}$ is the backward geometric mapping, as shown in Fig. \ref{Fig:2}.
Let $W = [w_{s,t}]$ be the involved interpolation kernel, the $y_{u,v}$ can be determined by the weighted sum of the pixel values defined in the closest neighbor $P = [x_{i,j}]$.
In implementation, the $W$ can usually be factorized into $W^{(V)}$ and  $W^{(H)}$, along the vertical and horizontal directions, i.e., $W = W^{(V)} \cdot (W^{(H)})^{\tau}$, thus we have, 
\begin{equation} \label{interpolation}
y_{u,v} = round((W^{(V)})^{\tau} \cdot P \cdot W^{(H)}),
\end{equation}
where $round(\cdot)$ is the rounding function, $W^{(V)}$ and $W^{(H)}$ are the 1-D interpolation kernels in column vector.

For pre-filtering based anti-aliasing scaling, the high frequency components of the original image $X$ are attenuated with an anti-aliasing filter $W_{anti}$ to avoid aliasing before sampling at pixel rates.
With an interpolation kernel $W_{s}$ in standard form (bilinear and bicubic), the equivalent anti-aliasing interpolation kernel can be re-written as,
\begin{equation} \label{anti-interpolation}
W = W_{anti} \cdot W_{s}.
\end{equation}

Generally speaking, the interpolation kernel has bell-shaped distribution and its size varies with the SFs. 
The smaller the SFs, the larger the kernel size becomes. Take the anti-aliasing bilinear interpolation scaling at $S\!F = 0.25$ and 0.5 as example, the corresponding sizes for kernel $W$ or interpolation block $P$ are $8 \times 8$ and $4 \times 4$, respectively, as compared to the fixed block of $2 \times 2$ for interpolation in standard form. 

\subsection{Degree of Pixel Involvement}
In the proposed framework of robust image steganography against general scaling attacks, the key idea is to seek the proxy stego image $X^{\prime} = X + \Delta\!X$ according to cover $X$, pre-scaled stego $Y^{\prime}$, the involved interpolation method and its SFs by incorporating the constrained integer programming. 
To identify the modifiable pixels ($\Delta\!x_{i,j} \geq 0$ or $\Delta\!x_{i,j} \leq 0$) in cover image, a metric known as the degree of pixel involvement (dPI) is introduced.
For a pixel $x_{i,j}$ in $X$, its dPI is defined as the number of its involvement in the interpolation computation of embeddable pixels in pre-scaled cover image $Y$.
To allow pixels in $X$ with relatively larger dPI (e.g., ${\rm dPI}(x_{i,j}) \geq 2$) to be modifiable pixels to generate the proxy stego image $X^{\prime}$ would lead to either the task of constrained integer programming for inverse interpolation hard to be solved, or the variations of modifiable pixels in $X^{\prime}$, in response to the embedding changes in scaled stego image $Y^{\prime}$, extremely large as compared to cover image $X$.

\subsection{Related Works}
At present, comprehensive research in depth on the robust image steganography capable of resisting to general scaling attacks at arbitrary SFs is not yet available on record. 
Recently, Zhu et al. \cite{zhang2018image} proposed a robust steganographic scheme, that can effectively resist the scaling attack with nearest neighbor interpolation, by incorporating the invariant pixels.
The most relevant work to our proposed method is the one in \cite{zhu2021inverse}, where the scheme in \cite{zhang2018image} is generalized to the scaling attacks with anti-aliasing bilinear (for SFs in $(0,0.5]$) and bicubic (for SFs in $(0,0.25]$) interpolations by exploring the inverse interpolation, although it could not yet be used directly for scaling attacks with the same interpolation methods in standard form.

\begin{figure*}[t]
\centering
\subfigure[Zhu's Scheme]{\label{Fig:3a}
\includegraphics[width=0.45\linewidth]{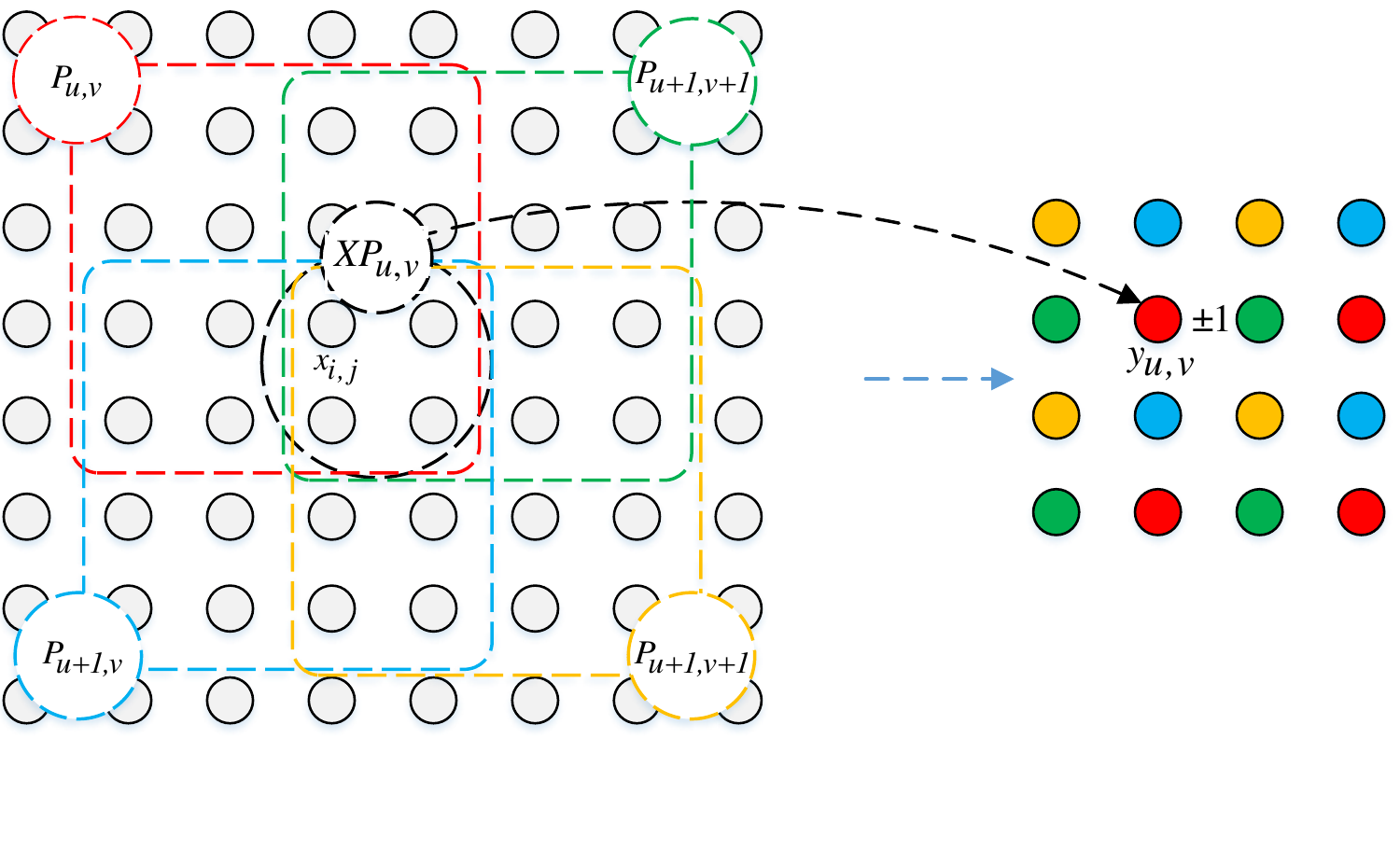}}
\subfigure[Our Scheme]{\label{Fig:3b}
\includegraphics[width=0.45\linewidth]{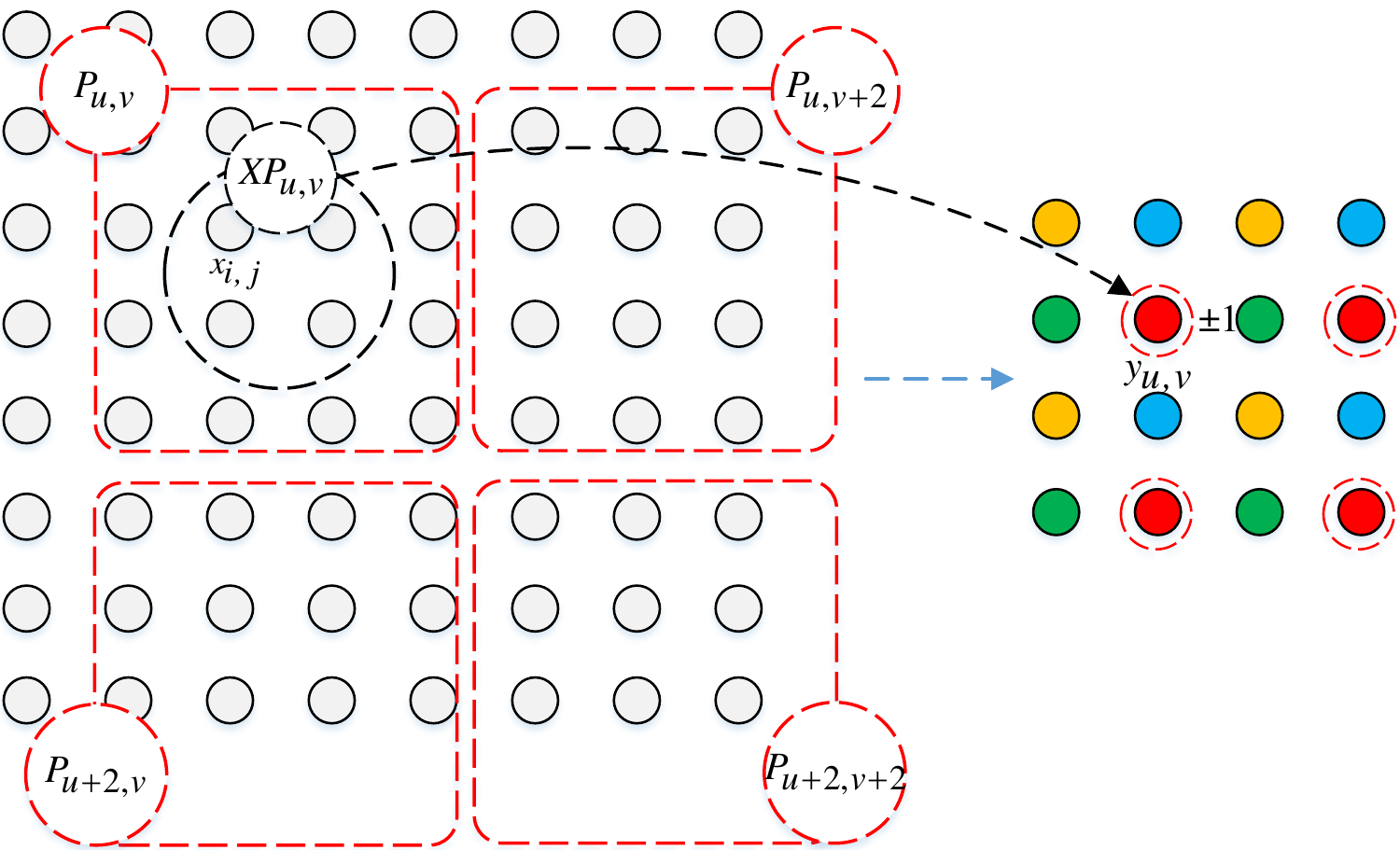}}\\
\centering
\caption{The diagram of robust image steganography schemes for anti-aliasing bilinear interpolation with $S\!F = 0.5$.}
\label{Fig:3}
\end{figure*}

In order to more clearly illustrate Zhu's method \cite{zhu2021inverse} as compared to our approach, we take the anti-aliasing bilinear interpolation at $S\!F = 0.5$ as an example. 
In Zhu's method, each pixel in the scaled cover image is assigned to be the embeddable one, consequently, the dPI value for most of the pixels in the cover image is $4$ except for a few pixels on the boundary of the image.
Let's denote the $4$ adjacent embeddable pixels in scaled image $Y$ as $y_{u,v}, \cdots, y_{u+1,v+1}$ and their corresponding closest neighbors or interpolation blocks of $4 \times 4$ in cover image $X$ as $P_{u,v}, \cdots, P_{u+1,v+1}$, as shown in Fig. \ref{Fig:3a}. 
Zhu's method tries to find the intersection block of $2 \times 2$ among $P_{u,v}, \cdots, P_{u+1,v+1}$ in $X$, i.e., $X\!P_{u,v}$ bounded by the black round box as shown in Fig.3(a), which we call supporting block in $X$ for  $y_{u,v}$ in the process of inverse interpolation. 
Note that all the 4 pixels in $X\!P_{u,v}$ involve in the interpolation computation of $y_{u, v}, y_{u, v+1}, y_{u+1, v}$, and $y_{u+1, v+1}$ at the same time, which is always achievable for anti-aliasing bilinear interpolation for $S\!F < 0.5$. 
Following the notion in (\ref{interpolation}), let $W_{u,v} = W_{u,v}^{(V)}(W_{u,v}^{(H)})^{\tau}, \cdots, W_{u+1,v+1} = W_{u+1,v+1}^{(V)}(W_{u+1,v+1}^{(H)})^{\tau}$ to be the partial weighs for $X\!P_{u,v}$ corresponding to $y_{u, v}, \cdots, y_{u+1, v+1}$, respectively, the inverse interpolation equation set for $\Delta\!X\!P_{u,v}$ is built by assigning the $X\!P_{u,v}$  is only contributed to $y_{u,v} \pm 1$ for the possible embedding changes at $y_{u, v}, \cdots, y_{u+1, v+1}$, i.e.,
\begin{equation} \label{Zhu'scheme}
\left\{\begin{array}{l}
round((W_{u,v}^{(V)})^{\tau} \cdot \Delta\!X\!P\cdot W_{u,v}^{(H)})~=\pm 1\\
round((W_{u,v+1}^{(V)})^{\tau} \cdot \Delta\!X\!P \cdot W_{u,v+1}^{(H)})~=~0\\
round((W_{u+1,v}^{(V)})^{\tau} \cdot \Delta\!X\!P \cdot W_{u+1,v}^{(H)})~=~0\\
round((W_{u+1,v+1}^{(V)})^{\tau} \cdot \Delta\!X\!P \cdot W_{u+1,v+1}^{(H)})~=~0\\
\end{array},
\right.
\end{equation}
where $\Delta\!X\!P_{u,v}$ is the variations compared to $X\!P_{u,v}$ in response to the embedding at $y_{u,v}$ in the inverse interpolation. 
The problems with the method in \cite{zhu2021inverse} are two-fold: (1) for given $y_{u,v}$ in $Y$, the determined $X\!P_{u,v}$ is inevitably located in the boundary of $P_{u,v}$ with smaller weights, leading to relatively larger modification $\Delta\!X\!P_{u,v}$ due to the embedding change at $y_{u,v}$; (2) for scaling attacks with anti-aliasing bilinear interpolation at $S\!F > 0.5$ (or bicubic at $S\!F > 0.25$), the dPI values for most of pixels in cover image $X$ are greater than $4$, and the established constraint equation sets tend to be over-determined accordingly, which are usually unsolvable. 
These issues, however, could be effectively solved with our proposed method, which will be elaborated later in Section \ref{ourframework}. 

\section{The Framework of Robust Image Steganography Against General Interpolation Scaling Attacks} \label{ProposedMethod}
In this Section, the framework of robust image steganography against general interpolation scaling attacks is proposed, which includes a tractable solver for the constrained integer programming corresponding to inverse interpolation, and a customized design of embedding distortion function for scaled images. 
\subsection{Image Inverse Interpolation As Constrained Integer Programming} \label{ourframework}
The proposed framework for robust image steganography against known general scaling attacks is shown in Fig. \ref{Fig:1}, where the proxy stego image $X^{\prime} = X + \Delta\!X$ with the same dimension as cover $X$ is generated from pre-scaled stego $Y^{\prime}$ through inverse interpolation. 
And the task of robust image steganography can then be formulated as the one of constrained integer programming with the objective to obtain the proxy stego image $X^{\prime}$, which could not only survive the general interpolation scaling attacks, but also be statistically indistinguishable from cover image $X$. 

For each embeddable pixel $y_{u,v}$ in pre-scaled stego $Y^{\prime}$, let $P_{u,v}$ of size $p \times p$ and $X\!P_{u,v}$ be its interpolation and supporting blocks in  $X$, the inverse interpolation is utilized by taking advantage of the constraint between $y_{u,v}$ and its supporting block $X\!P_{u,v}$ to obtain the variation $\Delta\!X\!P_{u,v}$ due to the embedding change at $y_{u,v}$. 
Unlike Zhu's method \cite{zhu2021inverse}, where the involved supporting blocks are composed of $2 \times 2$ pixels with ${\rm dPI} = 4$, the proposed method assigns a unique supporting block $X\!P_{u,v}$ of $2 \times 2$  adjacent pixels centered on its interpolation block $P_{u,v}$ in $X$ for each embeddable pixel $y_{u,v}$ in pre-scaled stego $Y^{\prime}$ as shown in Fig. \ref{Fig:3b}.
Note that the $X\!P_{u,v}$ in our method includes sufficient pixels with ${\rm dPI} = 1$ to be modified in inverse interpolation corresponding to the embedding change at $y_{u,v}$. 
This allocation of supporting blocks could ensure the task of inverse interpolation is always solvable and maintain the less variations for $\Delta\!X$ in inverse interpolation to improve the security performance of proxy stego image $X^{\prime}$.
To this end, the scaled image $Y$ is split into the sub-images $Y\!E\!P$ and $Y\!I\!P$, which are composed of embeddable pixels and idle pixels, respectively.
For a given interpolation method with $S\!F$ and interpolation block $P$ of $p \times p$, the $Y\!E\!P$ for data embedding is generated by sampling the scaled image $Y$ with interval $s$ along the horizontal and vertical directions. 
The reference sampling intervals $s$ for different interpolation methods with various $SFs$ are summarized in Table \ref{tab:samplestep}, which is a payload first design.
It is observed that, for std bilinear interpolation with $SF$ in $(0,0.5]$ and std bicubic interpolation with $SF$ in $(0,0.25]$, the sampling interval $s = 1$ is adopted, i.e, the generated $Y\!E\!P$ is $Y$ itself, while for anti-aliasing bilinear and bicubic interpolations, $s = \lceil p/2 \rceil$.  
It is obvious that the effective payload for proxy stego image $X^{\prime}$ heavily depends on the size of the generated $Y\!E\!P$.
Take the anti-aliasing interpolation scaling attacks as an example, the sampling interval $s$ is decreased with the increase of $S\!F$.
For anti-aliasing scaling with relatively larger $S\!F$,  sufficient pixels (embeddable) in $Y$ are usually preserved for data embedding, which is quite affordable for practical applications. 

Recall that, in our proposed framework, for each embeddable pixel $y_{u,v}$ in $Y$, only the pixels with ${\rm dPI} = 1$ in its supporting block $X\!P_{u,v}$ rather than its interpolation block $P_{u,v}$ in $X$ are considered to be modified in inverse interpolation.
In general, allowing more pixels in $X\!P_{u,v}$ to be modified would lead to more subtle changes of $\Delta\!X\!P_{u,v}$, thus more statistically undetectable for the generated proxy stego image $X^{\prime}$.
And the number of pixels with ${\rm dPI} = 1$ in $X\!P_{u,v}$  tends to be decreased with the increase of $S\!F$ as shown in Table \ref{tab:samplestep}.
For practical applications, the determination of modifiable pixels in $X\!P_{u,v}$ are also subjected to other two issues besides their ${\rm dPI}$ values, i.e., the variation bound and over/underflow in reverse interpolation. 
In specific, let $M_{u,v} = [m_{i,j}]_{2 \times 2}$ be the mask matrix to identify the modifiable pixels in $X\!P_{u,v}$, i.e, $m_{i,j} = 1$ or $0$ when the pixel $x_{i,j}$ is modifiable or not. 
We then have the constraint equation for $\Delta\!X\!P$, i.e, the variation of supporting block $X\!P_{u,v}$ due to the embedding change $\Delta\!y_{u,v}$ at $y_{u,v}$,
\begin{equation} \label{Constrain}
round((W_{u,v}^{(V)})^{\tau}\!\cdot\!(M_{u,v}\!\odot\!\Delta\!X\!P_{u,v})\!\cdot\!W_{u,v}^{(H)}) = \Delta y_{u,v},
\end{equation}
where $W_{u,v}^{(V)}$ and $W_{u,v}^{(H)}$ are the partial interpolation weights for $X\!P_{u,v}$, $\odot$ is the operator for element-wise multiplication and $\Delta y_{u,v} \in \{0,1,-1\}$. 
For $\forall\!y_{u,v}\!\in\!Y\!E\!P$, let $X\!P_{u,v}^{s}$ be the possibly largest sub-set of $X\!P_{u,v}$, which consists of the modifiable pixels $x_{i,j}$ with associated weight $w_{i,j}$, and $\omega_{u,v} = \sum\limits_{i,j}w_{i,j}$ for $x_{i,j}\!\in\!X\!P_{u,v}^{s}$. 
For $\Delta y_{u,v}\!=\!0$, a trivial solution for (\ref{Constrain}) is readily available, i.e., $\Delta\!X\!P_{u,v}\!=\!0$, while for $\Delta\!y_{u,v}\!=\!\pm\!1$, the variation bound for $\forall\!y_{u,v}\!\in\!Y\!E\!P$ can be determined as,
\begin{equation} \label{relevant}
\Delta_{u,v} = \lceil 1/\omega_{u,v} \rceil.
\end{equation}
The bound $\Delta_{u,v}$ is closely relevant to the security performance of the generated stego image $X^{\prime}$ and should be carefully determined.
Although the $\Delta_{u,v}$ is $X\!P_{u,v}$ dependent, according to our experimental results, in most cases, $\omega_{u,v} \geq 0.5$, and we could usually take $\Delta_{u,v} \leq 2$ (the possible maximum variations for $x_{i,j}$ are $\pm\!2$) in our implementation for security concerns. 
With given $\Delta_{u,v}$ and 8 bit grey scale image, we further let $x_{i,j} \pm \Delta_{u,v} \in \mathbb{N}_{256}$ to prevent the generated stego image $X^{\prime}$ from over/underflowing. 
Therefore, the sub-set $X\!P_{u,v}^{s}$ for $X\!P_{u,v}$ can be explicitly defined as,
\begin{equation} \label{sub-setXP}
\begin{array}{c}
X\!P_{u,v}^{s}\!=\!\{x_{i,j}|x_{i,j}\!\in\!X\!P_{u,v},{\rm dPI}(x_{i,j})\!=\!1,x_{i,j}\!\pm\!\Delta_{u,v}\!\in\!\mathbb{N}_{256}\\
~~~~~~~~~~~~~~~~~~~~~~~~~~~~~~~~~~~~~~~~~~{\rm and}~\lceil\!1\!/\!\sum\limits_{i,j}\!w_{i,j}\!\rceil\!\leq\!\Delta_{u,v}\}.
\end{array}
\end{equation}
And the $M_{u,v}$ for $X\!P_{u,v}$ can then be determined accordingly.
In specific, for $\forall x_{i,j} \in X\!P_{u,v}$, if $x_{i,j} \in X\!P_{u,v}^{s}$ then $m_{i,j} = 1$, otherwise $m_{i,j} = 0$.

\begin{table*}
\centering
\caption{The reference design parameters with the proposed method for various scaling channels and SFs, which consist of the size of the interpolation block $p \times p$, the sampling interval $s$, the number of pixels $N$ with ${\rm dPI}=1$ in the supporting box, and the achievable embedding rate $l$ in terms of the pre-scaled image.}
\label{tab:samplestep}
\scalebox{1}{
\begin{tabular}{lcccccccccccccccc}
\hline
\toprule [1 pt]
\multirow{2}{*}{\diagbox{${\rm SFs}$}{Channel}}&\multicolumn{4}{c}{Std bilinear}&\multicolumn{4}{c}{Std bicubic}&\multicolumn{4}{c}{Anti-aliasing bilinear}&\multicolumn{4}{c}{Anti-aliasing bicubic}\\
\cmidrule(r){2-5} \cmidrule(r){6-9} \cmidrule(r){10-13} \cmidrule(r){14-17}
&$p \times p$&$s$&$N$&$l$&$p \times p$&$s$&$N$&$l$&$p \times p$&$s$&$N$&$l$&$p \times p$&$s$&$N$&$l$\\
\hline
0.1&$2 \times 2$&1&4&$\sqrt{3}$&$4 \times 4$&1&4&$\sqrt{3}$&$20 \times 20$&10&4&$\sqrt{3}/100$&$40 \times 40$&20&4&$\sqrt{3}/400$\\
0.2&$2 \times 2$&1&4&$\sqrt{3}$&$4 \times 4$&1&4&$\sqrt{3}$&$9 \times 9$&5&4&$\sqrt{3}/16$&$17 \times 17$&9&4&$\sqrt{3}/81$\\
0.25&$2 \times 2$&1&4&$\sqrt{3}$&$4 \times 4$&1&4&$\sqrt{3}$&$8 \times 8$&4&4&$\sqrt{3}/16$&$16 \times 16$&8&4&$\sqrt{3}/64$\\
0.3&$2 \times 2$&1&4&$\sqrt{3}$&$4 \times 4$&2&4&$\sqrt{3}/4$&$7 \times 7$&4&4&$\sqrt{3}/16$&$14 \times 14$&7&4&$\sqrt{3}/49$\\
0.4&$2 \times 2$&1&4&$\sqrt{3}$&$4 \times 4$&2&4&$\sqrt{3}/4$&$5 \times 5$&3&4&$\sqrt{3}/9$&$10 \times 10$&5&4&$\sqrt{3}/25$\\
0.5&$2 \times 2$&1&4&$\sqrt{3}$&$4 \times 4$&2&4&$\sqrt{3}/4$&$4 \times 4$&2&4&$\sqrt{3}/4$&$8 \times 8$&4&4&$\sqrt{3}/16$\\
0.6&$2 \times 2$&2&4&$\sqrt{3}/4$&$4 \times 4$&2&4&$\sqrt{3}/4$&$5 \times 5$&3&4&$\sqrt{3}/9$&$7 \times 7$&4&4&$\sqrt{3}/16$\\
0.7&$2 \times 2$&2&4&$\sqrt{3}/4$&$4 \times 4$&2&4&$\sqrt{3}/4$&$3 \times 3$&2&4&$\sqrt{3}/4$&$6 \times 6$&3&2&$\sqrt{3}/9$\\
0.75&$2 \times 2$&2&4&$\sqrt{3}/4$&$4 \times 4$&3&4&$\sqrt{3}/9$&$3 \times 3$&2&2&$\sqrt{3}/4$&$6 \times 6$&3&2&$\sqrt{3}/9$\\
0.8&$2 \times 2$&2&4&$\sqrt{3}/4$&$4 \times 4$&3&4&$\sqrt{3}/9$&$3 \times 3$&2&1&$\sqrt{3}/4$&$5 \times 5$&3&2&$\sqrt{3}/9$\\
0.9&$2 \times 2$&2&4&$\sqrt{3}/4$&$4 \times 4$&3&4&$\sqrt{3}/9$&$3 \times 3$&2&1&$\sqrt{3}/4$&$5 \times 5$&3&1&$\sqrt{3}/9$\\
\hline
\toprule [1 pt]
\end{tabular}}
\end{table*}

We then proceed to the generation of the proxy stego image $X^{\prime} = X + \Delta\!X$ based on the cover image $X$, the involved interpolation scaling method with known $S\!F$ and the given payload of length $msg$. 
In the proposed method, the payload $msg$ is embedded into the embeddable sub-image $Y\!E\!P$ of the pre-scaled image $Y$ to obtain $Y\!E\!P^{\prime} = Emb(Y\!E\!P, \rho^{(Y)}, msg) = [y_{u,v} + \Delta\!y_{u,v}]$ by incorporating some existing SOTA steganographic schemes, e.g., S-UNIWARD \cite{holub2012designing} and HiLL \cite{li2014new}, with customized design for scaled images, which will be discussed later in Section \ref{DistortionFunction}.
On the other hand, for $\forall y_{u,v}\in Y\!E\!P$, there exists a unique correspondence between $y_{u,v}$ and its supporting block $X\!P_{u,v}$ in $X$. 
Denote $X^{(1)} = \mathop{\cup}\limits_{u,v}X\!P_{u,v}$, the cover image $X$ can then be decomposed into $X^{(1)}$ and its complement $X^{(2)}$.
The task of robust image steganography against scaling attacks amounts to determine the possible optimal variations   $\Delta\!X = \Delta\!X^{(1)} \cup \Delta\!X^{(2)}$ due to data embedding, which can be formulated as the following constrained integer programming (IP): 
\\
\\
\begin{equation} \label{IntegerObjective}
\mathop{\rm Min}\limits_{\Delta X} \,\, \Vert X - X^{\prime} \Vert_1,
\nonumber 
\end{equation}
\begin{equation} \label{IntegerConstrains}
\rm{s.t.} \left\{\begin{array}{l}
round((W_{u,v}^{(V)})^{\tau}\!\cdot\!(M_{u,v}\!\odot\!\Delta\!X\!P_{u,v})\!\cdot\!W_{u,v}^{(H)}) = \Delta\!y_{u,v},\\
~~~~~~~~~~~~~~~~~~~~~~~~~\Delta\!y_{u,v}\in\{0,\pm 1\}, \forall y_{u,v}\in Y\!E\!P\\
Y\!E\!P^{\prime} = [y_{u,v} + \Delta y_{u,v}] = Emb(\!Y\!E\!P, \rho^{(Y)}, msg)\\
X^{\prime} = X + \Delta X\\
X^{(1)} = \mathop{\cup}\limits_{u,v}X\!P_{u,v}\\
X = X^{(1)} \cup X^{(2)}\\
Y =  Y\!E\!P \cup Y\!I\!P
\end{array}\!,\!
\right.
\end{equation}
where $\rho^{(Y)}$ is the specifically designed embedding cost function for scaled images. 
Consequently, we have $\Delta\!X = \Delta\!X^{(1)} \cup \Delta\!X^{(2)}$, among which, the solution for $\Delta\!X^{(2)}$ is trivial and readily available, i.e., $\Delta\!X^{(2)} = 0$, while the one for $\Delta\!X^{(1)}$ is non-trivial and could be sought by incorporating the branch and bound algorithm (B\&B). 
The B\&B recursively splits the admissible solution space into a series of sub-solution spaces, i.e., branching, and searches the solution therein. 
To improve the efficiency of solution searching, the B\&B then keeps track of bounds on the minimum that is trying to find, and use the bounds to prune the search space to find the optimal solution, i.e., bounding. 
To apply B\&B to solve the constrained IP in (\ref{IntegerConstrains}), for $\forall y_{u,v}\in Y\!E\!P$, the searching for solution $\Delta X\!P_{u,v}$ could be classified into two scenarios according to if there exists a nonempty set $X\!P_{u,v}^{s}$ defined in (\ref{sub-setXP})  for $\Delta\!y_{u,v} \pm 1$, which are discussed as follows:
\begin{figure*}[t]
\centering
\centerline{\epsfig{figure=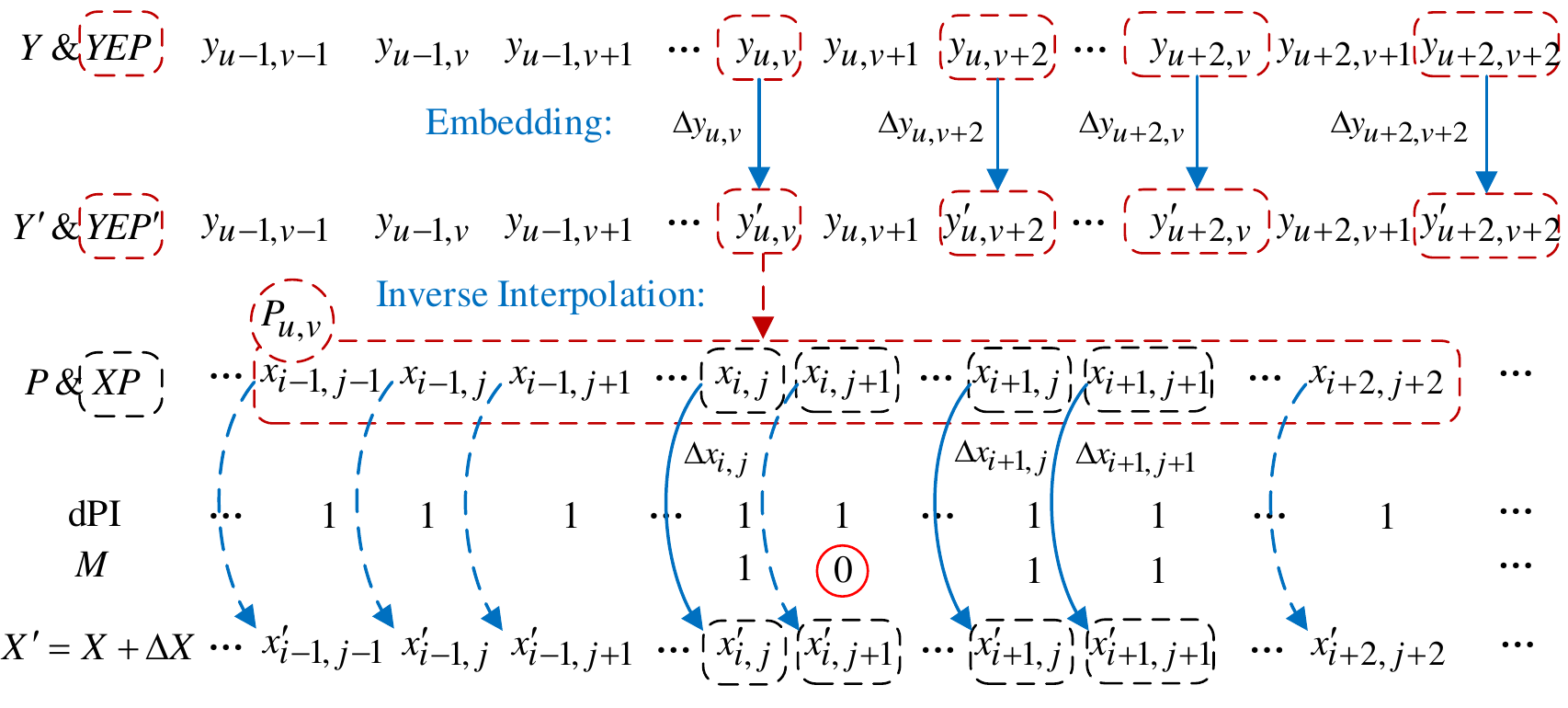,width=1\linewidth}}
\caption{The diagram of the Branch and Bound algorithm for bilinear interpolation scaling.} 
\label{Fig:4}
\end{figure*}
\begin{itemize}
    \item[1)] \textbf{The solution for $\Delta\!X\!P_{u,v}$ when $X\!P_{u,v}^{s} \neq \varnothing$}\\
    If $X\!P_{u,v}^{s} \neq \varnothing$ for $\Delta\!y_{u,v}= \pm 1$, the corresponding $y_{u,v}$ in $Y\!E\!P$ is an effective embeddable pixel, and the ternary embedding $\Delta\!y_{u,v}$ is allowed. 
    In fact, for practical applications, the vast majority of $\Delta\!X\!P_{u,v}$s in cover $X$ have their non-empty $X\!P_{u,v}^{s}$s for $\Delta\!y_{u,v}= \pm 1$. 
    When $\Delta\!y_{u,v} = 0$, the trivial solution $\Delta\!X\!P_{u,v} = 0$ is obtained, i.e., none of the pixels in $X\!P_{u,v}$ should be modified. 
    While for $\Delta\!y_{u,v}= \pm 1$, the solution space $\Delta\!X\!P_{u,v}$ is split into  $\Delta\!X\!P_{u,v}^{s}$ and its complement $\Delta\!X\!P_{u,v}^{c}$ by taking into the ${\rm dPI}$ value and over/underflowing. 
    Note that the weight for $\forall x_{i,j} \in X\!P_{u,v}$ is usually positive as $X\!P_{u,v}$ is centered on the interpolation block $P_{u,v}$. 
    Therefore, the strategy of unidirectional searching could be utilized. 
    Without loss of generality, we take $\Delta\!y_{u,v} = +1$ as an example to illustrate the solution searching process, the same applies to $\Delta\!y_{u,v} = -1$. 
    Assume $\Delta\!y_{u,v} = +1$, the $\Delta\!x_{i,j}$s in $\Delta\!X\!P_{u,v}^{s}$ are sorted in descending order in terms of their weights $w_{i,j}$s, and initialized as 0. 
    The  $\Delta\!x_{i,j}$s in $\Delta\!X\!P_{u,v}^{s}$ are then increased in an element-wise manner with step 1 within the variation bound $\Delta_{u,v}$ until the constraint, i.e., $round((W_{u,v}^{(V)})^{\tau}\!\cdot\!(M_{u,v}\!\odot\!\Delta\!X\!P_{u,v})\!\cdot\!W_{u,v}^{(H)}) = \Delta\!y_{u,v}$ is met. 
    For $\Delta\!x_{i,j}\in \Delta\!X\!P_{u,v}^{c}$, the solution is readily available, i.e., $\Delta\!X\!P_{u,v}^{c} = 0$.
    Fig. \ref{Fig:4} shows the diagram of how the proxy stego image $X^{\prime}$ is obtained from pre-scaled stego image $Y^{\prime}$ using IP-based inverse interpolation with B\&B search for anti-aliasing bilinear scaling with $S\!F = 0.5$. 
    For $y_{u,v} \in Y\!E\!P$ and $\Delta\!y_{u,v} = +1$, it is observed that, for the given variation bound $\Delta_{u,v}$, among the 4 pixels with ${\rm dPI} = 1$ in the corresponding supporting block $X\!P_{u,v} = \{x_{i,j}, \cdots, x_{i+1,j+1}\}$, only $x_{i,j}, x_{i+1,j}$, and $x_{i+1,j+1} \in X\!P_{u,v}^{s}$ are modifiable pixels in reverse interpolation, while $x_{i,j+1} \in X\!P_{u,v}^{c}$ is not allowed to be modified to prevent overflowing , as shown in Fig. \ref{Fig:4}. 
    \item[2)] \textbf{The solution for $\Delta\!X\!P_{u,v}$ when $X\!P_{u,v}^{s} = \varnothing$}\\
    If no $X\!P_{u,v}^{s}$ exist in $X\!P_{u,v}$, the embedding change $\Delta\!y_{u,v} = \pm 1$ would lead to the over/underflowing in reverse interpolation for all pixels with ${\rm dPI} = 1$ in the associated supporting block $X\!P_{u,v}$, as a result, the corresponding $y_{u,v}$ in $Y\!E\!P$ should be identified as a wet pixel by assigned an embedding cost of infinity for $\Delta\!y_{u,v} = \pm 1$, and $\Delta\!x_{i,j} = 0$ for $\forall x_{i,j} \in X\!P_{u,v}$.
\end{itemize}

Note that the proposed method could be easily generalized to the scaling channel with nearest neighbor interpolation, which is nothing more than a special case of the scaling channels under consideration. Under the circumstance, the embeddable sub-image $Y\!E\!P$ is the pre-scaled image $Y$ itself, and for $\forall y_{u,v} \in Y\!E\!P$, its interpolation block $P_{u,v}$ comprises of $2\times2$ pixels, in which the pixel nearest to $T^{-1}(u,v)$ is used to constitute the supporting block $X\!P_{u,v}$ of size 1x1 with ${\rm weight}=1$ and ${\rm dPI}=1$, where the $T^{-1}(u,v)$ is the backward mapping of coordinate $(u,v)$ in cover image $X$. As a summary, we also give the pseudo-code for the generation of the proxy stego image $X^{\prime}$ using IP-based inverse interpolation with B\&B search as shown in Algorithm \ref{alg:algorithm1}.

\begin{algorithm}[h]
	\caption{The IP-based inverse interpolation with B\&B to obtain  the proxy stego image $X^{\prime}$.}
	\label{alg:algorithm1}
	\KwIn{$X$, $Y$, and $Y^{\prime}$}
	\KwOut{$X^{\prime}$}  
	\BlankLine	
        Determine the design parameters according to Table \ref{tab:samplestep}, i.e, the sampling interval $s$, and obtain the $Y\!E\!P$;\\
	For $\forall y_{u,v} \in Y\!E\!P$, calculate the embedding change $\Delta\!y_{u,v}$ and identify its supporting block $X\!P_{u,v}$ in $X$, let $X^{(1)} = \mathop{\cup}\limits_{u,v}X\!P_{u,v}$;\\
        Decompose the $X$ into $X^{(1)}$ and its complement $X^{(2)}$, and initialize $\Delta\!X^{(1)}$ and $\Delta\!X^{(2)}$ with 0;\\        
	\For{{\rm each} $y_{u,v} \in Y\!E\!P$}{
	Determine the largest sub-set $X\!P^{s}_{u,v}$ in $X\!P_{u,v}$, its set size $N_{u,v}$ and the mask $M_{u,v}$ for $X\!P_{u,v}$ according to (\ref{sub-setXP});\\   	
	\If{$X\!P_{u,v}^{s} = \varnothing$}{$\Delta\!x_{i,j}=0$ for $\forall x_{i,j} \in X\!P_{u,v}$}
	\If{$X\!P_{u,v}^{s} \neq \varnothing \& \Delta\!y_{u,v} = 0$}{$\Delta\!x_{i,j}=0$ for $\forall x_{i,j} \in X\!P_{u,v}$}
	\If{$X\!P_{u,v}^{s} \neq \varnothing \& |\Delta\!y_{u,v}| = 1$}{\justify{Decompose the $\Delta\!X\!P_{u,v}$ into $\Delta\!X\!P_{u,v}^{s}$ and its complement $\Delta\!X\!P_{u,v}^{c}$, and assign $\Delta\!X\!P_{u,v}^{s} = 0$. The $\Delta\!x_{i,j}$s in $\Delta\!X\!P_{u,v}^{s}$ are initialized as 0, and sorted in descending order in terms of their weights $w_{i,j}$s. The coordinates of the sorted pixels in $\Delta\!X\!P_{u,v}^{s}$ are arranged as an array $Z[n]$ of length $N_{u,v}$;}\\
        $n= 0$;\\
	\While{$|round((W^{(V)}_{u,v})^{\tau}\!\cdot\!(M_{u,v}\!\odot\!X\!P_{u,v})\!\cdot\!W^{(H)}_{u,v})|\neq1$}{
	\If{$\Delta\!y_{u,v} = 1$}{$\Delta x_{Z[n]} = \Delta x_{Z[n]} + 1$;\\}
	\If{$\Delta\!y_{u,v} = -1$}{$\Delta x_{Z[n]} = \Delta x_{Z[n]} - 1$;\\}
	$n = n + 1$;\\
	\If{$n \geq N_{u,v}$}{$n = 0$;}}}
	}
	\textbf{return} $X^{\prime} = X + \Delta\!X$
\end{algorithm}

We then revisit the task of robust image steganography against anti-aliasing bilinear scaling attack with $S\!F=0.5$, as shown in Fig. \ref{Fig:3b}, to illustrate how the same problem is effectively solved within our proposed framework. 
For the given scaling channel, the convolution block $P_{u,v}$ is of size $4 \times 4$, and the sampling interval $s = 2$ is determined according to Table \ref{tab:samplestep}, which is then used to generate the embeddable sub-image $Y\!E\!P$ by down-sampling the pre-scaled image $Y$ of keeping one pixel out of two. 
The $Y\!E\!P$ is embedded with embedding rate $\alpha$ in terms of $Y\!E\!P$, or $\alpha/4$ in terms of $Y$, using one of existing SOTA steganographic schemes, e.g., S-UNIWARD or HiLL, to generate the $Y\!E\!P$. 
For the scaling attack under consideration, the supporting block $X\!P_{u,v}$ centered on interpolation block $P_{u,v}$ in $X$ associated with the embeddable pixel $y_{u,v}$ in $Y\!E\!P$ includes 4 pixels with ${\rm dPI} = 1$, which are modifiable with variation bound $\Delta_{u,v} = 2$.
By adopting the reference design parameters in Table \ref{tab:samplestep}, the variation $\Delta\!X$ due to the embedding change $\Delta\!Y\!E\!P$ is obtained through the IP-based inverse interpolation using Algorithm \ref{alg:algorithm1}, and the proxy stego image $Y^{\prime}$ is thus determined. 
Although the raw embedding capacity of the proposed method seems to be lower than Zhu's method \cite{zhu2021inverse}, the security performance of the generated proxy stego images with our method, however, is usually greatly superior to the one with Zhu's method at the same embedding rate, e.g.,  $\bar{P}_{E} = 0.24$ (the proposed) as compared to $\bar{P}_{E} = 0.02$ (the method in \cite{zhu2021inverse}) for the anti-aliasing bilinear scaling attack with $S\!F = 0.5$, for the same embedding rate $\alpha = 0.2 {\rm bpp}$ in terms of pre-scaled image $Y$. 
And the significantly improved ``secure'' capacity is usually more desirable in practical applications.

\subsection{The Construction of the Embedding Distortion Function for Scaled Images} \label{DistortionFunction}
In the proposed framework of robust image steganography against general scaling attacks, the secret data is embedded into the embeddable sub-image $Y\!E\!P$ of the scaled image $Y$, which is then used to generate the proxy stego image $X^{\prime}$ by IP-based inverse interpolation with B\&B algorithm. 
For $\forall y_{u,v} \in Y\!E\!P$, there exists a unique supporting block $X\!P_{u,v}$ composed of sufficient pixels $x_{i,j}$ with ${\rm dPI}(x_{i,j}) = 1$ in cover image $X$, and a mask matrix $M_{u,v} = [m_{i,j}]$ is adopted to identify which pixels in $X\!P_{u,v}$ are modifiable in the inverse interpolation. 
In general, the embedding distortion function of some existing SOTA steganographic schemes could be used to evaluate the cost for embeddable pixel $y_{u,v}$ based on the scaled sub-image $Y\!E\!P$, as it is done in \cite{zhu2021inverse}. 
Considering the fact that the objective of robust steganography against scaling attacks is to maintain the statistical indistinguishability of the proxy stego image $X^{\prime}$ from the cover image $X$, rather than the scaled stego image $Y^{\prime}$ from the scaled cover image $Y$. 
It is expected that the security performance of the proxy stego image $X^{\prime}$ would be improved if the embedding cost for the scaled image $Y$ is determined by incorporating the statistics of cover image $X$. 
We thus have two different ways of how to design embedding costs for scaled images, i.e., the ``plain'' approach based on the scaled images and the ``Pro'' approach based on the cover images. 
Let X be one of the existing SOTA steganographic methods, the convention X-Plain and X-Pro, e.g., S-UNIWARD-Plain and S-UNIWARD-Pro, are employed to denote the scaled image based and cover image based embedding cost design for scaled images, respectively. 

Let $\psi$ be the adopted distortion function for an existing SOTA steganographic method, for scaled image based approach, the embedding cost is directly derived from $Y\!E\!P$, i.e., the embeddable sub-image of scaled cover image $Y$: 
\begin{equation}
\left\{\begin{array}{l}
\rho^{({\rm Plain})(\pm 1)}_{u,v} = \psi^{\pm 1}_{u,v},~y_{u,v} \in  Y\!E\!P~{\rm and}~X\!P_{u,v}^{s} \neq \varnothing\\
\rho^{({\rm Plain})(\pm 1)}_{u,v} = +\infty,~y_{u,v} \in  Y\!E\!P~{\rm and}~X\!P_{u,v}^{s} = \varnothing\\
\rho^{({\rm Plain})(+0)}_{u,v} = 0,~~~~~y_{u,v} \in  Y\!E\!P\\
\end{array}\!,
\right.
\end{equation}
where $X\!P_{u,v}^{s} = \varnothing$ indicates that the embedding change $(\pm 1)$ at $y_{u,v}$ in $Y\!E\!P$ would inevitably lead to the under/overflowing for $\forall x_{i,j} \in X\!P_{u,v}$ and ${\rm dPI}(x_{i,j}) = 1$, and $y_{u,v}$ is identified as a wet pixel with the cost of infinity.

Alternatively, we have the customized design of embedding cost $\rho^{({\rm Pro})}$ for $Y\!E\!P$ by taking into account the statistics of cover image $X$, which is defined as:
\begin{equation}
\left\{\begin{array}{l}
\rho^{({\rm Pro})(\pm 1)}_{u,v} = \frac{\sum\limits_{i,j}m_{i,j}\cdot\psi_{i,j}^{(\pm 1)}}{\sum\limits_{i,j}m_{i,j}},~y_{u,v} \in  Y\!E\!P~{\rm and}~X\!P_{u,v}^{s} \neq \varnothing\\
\rho^{({\rm Pro})(\pm 1)}_{u,v} = +\infty,~y_{u,v} \in  Y\!E\!P~{\rm and}~X\!P_{u,v}^{s} = \varnothing\\
\rho^{({\rm Pro})(+0)}_{u,v} = 0,~~~~~y_{u,v} \in  Y\!E\!P\\
\end{array}\!,
\right.
\end{equation}
where the $\psi_{i,j}^{(\pm 1)}$ is the embedding cost for $x_{i,j} (x_{i,j} \in X\!P_{u,v})$ derived from the cover image, and $m_{i,j}$ is the mask, which is used to identify whether the corresponding pixel $x_{i,j}$ is actually involved in the reverse interpolation. 

\section{Experiment Results and Analysis}
In this Section, extensive experiments are carried out to demonstrate the effectiveness of the proposed robust image steganographic framework against general scaling attacks with standard bilinear and bicubic interpolations and their anti-aliasing variants.
We evaluate the feasibility of our method for various scaling channels and discuss its implementation with different design parameters for trade-off between security and payload, and compare our method with prior arts in terms of security performance for different interpolation scaling methods with various SFs and payloads using the SOTA steganalyzer SRM \cite{fridrich2012rich}.
\subsection{Experimental Settings}
In this paper, the involved experiments are carried out on BOSSbase 1.01 \cite{Bas2011Break}, which comprises of 10,000 $512 \times 512 \times 8$-bit gray-scale images with diverse texture characteristics, and is widely adopted in image steganography.  
For given interpolation scaling channel with a specific $S\!F$, the pre-scaled cover images $Y$s and their embeddable sub-images $Y\!E\!P$s are generated. 
It is worth noting that, throughout the paper, \textbf{the embedding rate $\alpha$ for proxy stego image $X^{\prime}$ is evaluated in terms of the pre-scaled image $Y$ rather than the $X^{\prime}$ itself}, and the achievable embedding rates for various scaling channels are summarized in Table \ref{tab:samplestep}. 
Note that, in the interest of possibly higher embedding rate, throughout the experimental section, unless otherwise specified, the embedding rate with the proposed scheme for various scaling channel is calculated according to the reference design parameters (the sampling interval $s$) in Table \ref{tab:samplestep}. 
Two SOTA adaptive image steganographic schemes in spatial domain, namely, S-UNIWARD (Spatial UNIversal WAvelet Relative Distortion) \cite{holub2012designing} and HiLL (High-pass, Low-pass and Low-pass) \cite{li2014new}, are explored to design the customized  distortion functions for scaled images  within the framework of minimal distortion embedding, either in ``Plain" or ``Pro" approaches. 
The secret data are then embedded into $Y\!E\!P$s using ternary syndrome-trellis code (STC) with parity-check matrix of $h=10$, which are then used to generate the proxy stego image $X^{\prime}$s through IP-based inverse interpolation.  
To evaluate the security performance (i.e, statistical undetectability) of the involved robust image steganographic methods, we have an image database with 10,000 pairs of cover and proxy stego images of size $512 \times 512 \times 8$-bit for a given embedding method, scaling channel and payload. 
The SOTA steganalyzer SRM-34,671D \cite{fridrich2012rich} with the Fisher linear discriminant ensemble classifier for spatial images is adopted.  
In our implementation, half of the cover and the proxy stego image pairs are randomly selected as the training set for the ensemble classifier, and the remaining half are used as test set to evaluate the trained classifier. 
The security performance is quantified as the minimal total probability of error under equal priors achieved on the test set by ten times of randomly testing, denoted as $\bar{P}_{E}$.

\subsection{The Comparison of Feasibility to Various Scaling Channels}
The objective of robust image steganography against scaling attacks is to survive the scaling channels while maintaining the acceptable security performance. 
From the perspective of communication, the feasibility of a robust steganographic scheme could be evaluated in terms of robustness by its capability to reliably recover the embedded data from the noisy channels with and without channel coding. 
To this end, the constraints between proxy stegos and the pre-scaled stegos due to embedding changes should be taken into account in reverse interpolation. 
As noted earlier, the proposed method exhibits much better applicability to various scaling channels by identifying the modifiable pixels in cover image with the ${\rm dPI}$ metric and thus efficiently solving the IP-based inverse interpolation with B\&B algorithm. 
The comparison of the proposed method with other SOTA methods in terms of feasibility is summarized in Table \ref{tab:robustness}, where ``\Checkmark'' and  ``\XSolidBrush'' stands for the associated method is and is not applicable to the corresponding scaling channel, respectively. 
What stands out in Table \ref{tab:robustness} is that the proposed method could perfectly recover the embedded data, even without channel coding, for all the involved scaling attacks with arbitrary SFs, while other competing methods could be only applicable to some specific scaling channels with relatively smaller SFs. 
To be specific, the method in \cite{zhang2018image} is specifically designed for scaling channel with nearest neighbor interpolation, and the one in \cite{zhu2021inverse} is for the scaling attacks with anti-aliasing bilinear ( for SFs in $(0,0.5]$ ) and bicubic (for SFs in $(0,0.25]$)  interpolation.
\begin{table*}[t]
\centering
\caption{THE COMPARISON OF THE PROPOSED METHOD WITH OTHER COMPETING METHODS IN TERMS OF FEASIBILITY TO VARIOUS SCALING CHANNELS.}
\label{tab:robustness}
\scalebox{1}{
\begin{tabular}{cccccccc}
\hline
\toprule [1 pt]
\multirow{2}{*}{\diagbox{Method}{Channel}}&\multirow{2}{*}{Nearest}&\multirow{2}{*}{Std Bilinear}&\multirow{2}{*}{Std bicubic}&\multicolumn{2}{c}{Anti-aliasing bilinear}&\multicolumn{2}{c}{Anti-aliasing bilinear}\\
\cmidrule(r){5-6} \cmidrule(r){7-8}
&$S\!F \in$(0,1)&$S\!F \in$(0,1)&$S\!F \in$(0,1)&$S\!F \in$(0,0.5]&$S\!F \in$(0.5,1)&$S\!F \in$(0,0.25]&$S\!F \in$(0.25,1)\\
\hline
\cite{zhang2018image}&\Checkmark &\XSolidBrush &\XSolidBrush &\XSolidBrush & \XSolidBrush &\XSolidBrush &\XSolidBrush\\
\hline
\cite{zhu2021inverse}&\XSolidBrush &\XSolidBrush &\XSolidBrush &\Checkmark & \XSolidBrush &\XSolidBrush &\XSolidBrush\\
\hline
Ours&\Checkmark &\Checkmark &\Checkmark &\Checkmark & \Checkmark &\Checkmark &\Checkmark\\
\hline
\toprule [1 pt]
\end{tabular}}
\end{table*}
\subsection{The Security Performance of The Proposed Method}
\begin{table*}[t]
\centering
\caption{THE SECURITY PERFORMANCE BETWEEN COVER AND PROXY STEGO IMAGES IN TERMS OF $\bar{P}_{E}$ WITH THE PROPOSED METHOD FOR SCALING CHANNELS WITH STD BILINEAR AND BICUBIC INTERPOLATIONS AT VARIOUS PAYLOADS (WITHIN THE ACHIEVABLE EMBEDDING RATES) AND SFs.}
\label{tab:security-std}
\scalebox{1}{
\begin{tabular}{cccccccccccc}
\hline
\toprule [1 pt]
\multirow{2}{*}{SFs}&\multirow{2}{*}{Distortion}&\multicolumn{5}{c}{Relative payload $\alpha$ (bpp) for std bilinear scaling}&\multicolumn{5}{c}{Relative payload $\alpha$ (bpp) for std bicubic scaling}\\
\cmidrule(r){3-7} \cmidrule(r){8-12}
&Function&0.025&0.05&0.1&0.2&0.3&0.025&0.05&0.1&0.2&0.3\\
\hline
\multirow{4}{*}{0.3}
&S-UNIWARD-Plain&0.4975&0.4951&0.4843&0.4583&0.4290&0.4966&0.4869&0.4599&0.3793&0.2868\\
&S-UNIWARD-Pro&\textbf{0.4980}&\textbf{0.4959}&\textbf{0.4871}&\textbf{0.4610}&\textbf{0.4337}&\textbf{0.4967}&\textbf{0.4887}&\textbf{0.4602}&\textbf{0.3865}&\textbf{0.3020}\\
&HiLL-Plain&0.4982&0.4954&0.4877&0.4635&0.4390&0.4971&0.4907&0.4653&0.3990&0.3041\\
&HiLL-Pro&\textbf{0.4991}&\textbf{0.4975}&\textbf{0.4910}&\textbf{0.4723}&\textbf{0.4550}&\textbf{0.4980}&\textbf{0.4922}&\textbf{0.4695}&\textbf{0.4086}&\textbf{0.3202}\\
\hline
\multirow{4}{*}{0.5}
&S-UNIWARD-Plain&0.4943&0.4832&0.4548&0.3930&0.3282&0.4857&0.4593&0.3930&0.2602&0.1417\\
&S-UNIWARD-Pro&\textbf{0.4945}&\textbf{0.4841}&\textbf{0.4580}&\textbf{0.3970}&\textbf{0.3404}&\textbf{0.4878}&\textbf{0.4633}&\textbf{0.3982}&\textbf{0.2718}&\textbf{0.1612}\\
&HiLL-Plain&0.4963&0.4878&0.4666&0.4059&0.3476&0.4902&0.4678&0.4135&0.2884&0.1669\\
&HiLL-Pro&\textbf{0.4972}&\textbf{0.4902}&\textbf{0.4724}&\textbf{0.4131}&\textbf{0.3589}&\textbf{0.4913}&\textbf{0.4705}&\textbf{0.4149}&\textbf{0.3013}&\textbf{0.1826}\\
\hline
\multirow{4}{*}{0.7}
&S-UNIWARD-Plain&0.4564&0.4112&0.3230&0.2023&0.1194&0.4778&0.4458&0.3732&0.2442&0.1316\\
&S-UNIWARD-Pro&\textbf{0.4626}&\textbf{0.4163}&\textbf{0.3340}&\textbf{0.2133}&\textbf{0.1270}&\textbf{0.4793}&\textbf{0.4491}&\textbf{0.3792}&\textbf{0.2454}&\textbf{0.1386}\\
&HiLL-Plain&0.4633&0.4219&0.3329&0.2061&0.1202&0.4860&0.4615&0.4042&0.2928&0.1704\\
&HiLL-Pro&\textbf{0.4677}&\textbf{0.4252}&\textbf{0.3388}&\textbf{0.2139}&\textbf{0.1285}&\textbf{0.4871}&\textbf{0.4625}&\textbf{0.4072}&\textbf{0.2955}&\textbf{0.1793}\\
\hline
\toprule [1 pt]
\end{tabular}}
\end{table*}

\begin{table*}[t]
\centering
\caption{THE SECURITY PERFORMANCE BETWEEN COVER AND PROXY STEGO IMAGES IN TERMS OF $\bar{P}_{E}$ WITH THE PROPOSED METHOD FOR SCALING CHANNELS WITH ANTI-ALIASING BILINEAR AND BICUBIC INTERPOLATIONS AT VARIOUS PAYLOADS (WITHIN THE ACHIEVABLE EMBEDDING RATES) AND SFs.}
\label{tab:security-antialiasing}
\scalebox{1}{
\begin{tabular}{cccccccccccccc}
\hline
\toprule [1 pt]
\multirow{2}{*}{SFs}&\multirow{2}{*}{Distortion}&\multicolumn{6}{c}{Relative payload $\alpha$ (bpp) for anti-aliasing bilinear scaling}&\multicolumn{6}{c}{Relative payload $\alpha$ (bpp) for anti-aliasing bicubic scaling}\\
\cmidrule(r){3-8} \cmidrule(r){9-14}
&Function&0.005&0.01&0.025&0.05&0.075&0.1&0.005&0.01&0.025&0.05&0.075&0.1\\
\hline
\multirow{4}{*}{0.3}
&S-UNIWARD-Plain&0.4723&0.4335&0.3318&0.2238&--&--&0.4887&0.4671&0.3970&--&--&--\\
&S-UNIWARD-Pro&\textbf{0.4781}&\textbf{0.4445}&\textbf{0.3517}&\textbf{0.2387}&--&--&\textbf{0.4906}&\textbf{0.4730}&\textbf{0.4038}&--&--&--\\
&HiLL-Plain&0.4766&0.4431&0.3589&0.2530&--&--&0.4888&0.4701&0.4092&--&--&--\\
&HiLL-Pro&\textbf{0.4850}&\textbf{0.4550}&\textbf{0.3788}&\textbf{0.2744}&--&--&\textbf{0.4932}&\textbf{0.4791}&\textbf{0.4221}&--&--&--\\
\hline
\multirow{4}{*}{0.5}
&S-UNIWARD-Plain&0.4974&0.4921&0.4655&0.4141&0.3700&0.3291&0.4914&0.4802&0.4280&0.3391&0.2607&--\\
&S-UNIWARD-Pro&\textbf{0.4986}&\textbf{0.4922}&\textbf{0.4715}&\textbf{0.4302}&\textbf{0.3889}&\textbf{0.3519}&\textbf{0.4968}&\textbf{0.4858}&\textbf{0.4348}&\textbf{0.3495}&\textbf{0.2717}&--\\
&HiLL-Plain&0.4980&0.4936&0.4756&0.4418&0.4068&0.3752&0.4965&0.4898&0.4504&0.3750&0.2974&--\\
&HiLL-Pro&\textbf{0.4983}&\textbf{0.4954}&\textbf{0.4808}&\textbf{0.4498}&\textbf{0.4203}&\textbf{0.3860}&\textbf{0.4970}&\textbf{0.4905}&\textbf{0.4545}&\textbf{0.3750}&\textbf{0.3055}&--\\
\hline
\multirow{4}{*}{0.7}
&S-UNIWARD-Plain&0.4915&0.4810&0.4229&0.3071&0.2533&0.1890&0.4962&0.4922&0.4722&0.4174&0.3532&0.2874\\
&S-UNIWARD-Pro&\textbf{0.4955}&\textbf{0.4872}&\textbf{0.4394}&\textbf{0.3393}&\textbf{0.2831}&\textbf{0.2269}&\textbf{0.4972}&\textbf{0.4940}&\textbf{0.4743}&\textbf{0.4215}&\textbf{0.3612}&\textbf{0.2966}\\
&HiLL-Plain&0.4968&0.4902&0.4603&0.3954&0.3343&0.2749&0.4988&0.4967&0.4793&0.4386&0.3871&0.3290\\
&HiLL-Pro&\textbf{0.4988}&\textbf{0.4936}&\textbf{0.4708}&\textbf{0.4185}&\textbf{0.3666}&\textbf{0.3123}&\textbf{0.4998}&\textbf{0.4979}&\textbf{0.4814}&\textbf{0.4427}&\textbf{0.3951}&\textbf{0.3382}\\
\hline
\toprule [1 pt]
\end{tabular}}
\end{table*}
\subsubsection{The Security Performance with Various Embedding Costs}
To verify the effectiveness of our customized embedding cost design for scaled images, we show the security performance with the proposed method for various scaling channels at different embedding rates. In specific, two SOTA steganographic schemes, e.g., S-UNIWARD and HiLL, equipped with two different customized embedding costs derived from scaling (plain) and cover images (Pro) respectively, are adopted.  
Table \ref{tab:security-std} and \ref{tab:security-antialiasing} shows the security performance with the proposed method (X-Plain and X-Pro) for various std and anti-aliasing interpolation scaling channels at different payloads. 
It is observed that, no matter whether they are std or anti-aliasing bilinear and bicubic scaling channels, the Pro-based approaches, i.e., S-UNIWARD-Pro and HiLL-Pro generally outperform their Plain-based counterparts, i.e., S-UNIWARD-Plain and HiLL-Plain, especially at relatively high embedding rates.  
For std bilinear interpolation scaling channels with $S\!F=0.5$, the performance of S-UNIWARD-Pro and HiLL-Pro are increased by 1.22\% and 1.13\% respectively at 0.3bbp, compared with S-UNIWARD-Plain and HiLL-Plain, as shown in Table \ref{tab:security-std}, while for anti-aliasing bilinear interpolation scaling channels with SF=0.5, the performance of S-UNIWARD-Pro and HiLL-Pro are also increased by 1.89\% and 1.35\% at 0.75bpp, compared with their Plain-based counterparts, as shown in Table \ref{tab:security-antialiasing}. 
As Table \ref{tab:security-std} and \ref{tab:security-antialiasing} show, the proposed scheme could be well applicable to robust image steganography against various scaling attacks with quite affordable payloads and acceptable security, and the Pro-based approaches equipped with the embedding costs derived from the cover images are more appropriate for robust steganographic communication over scaling channels by incorporating the statistics of cover images. 
\begin{figure}[t]
\centering
\centerline{\epsfig{figure=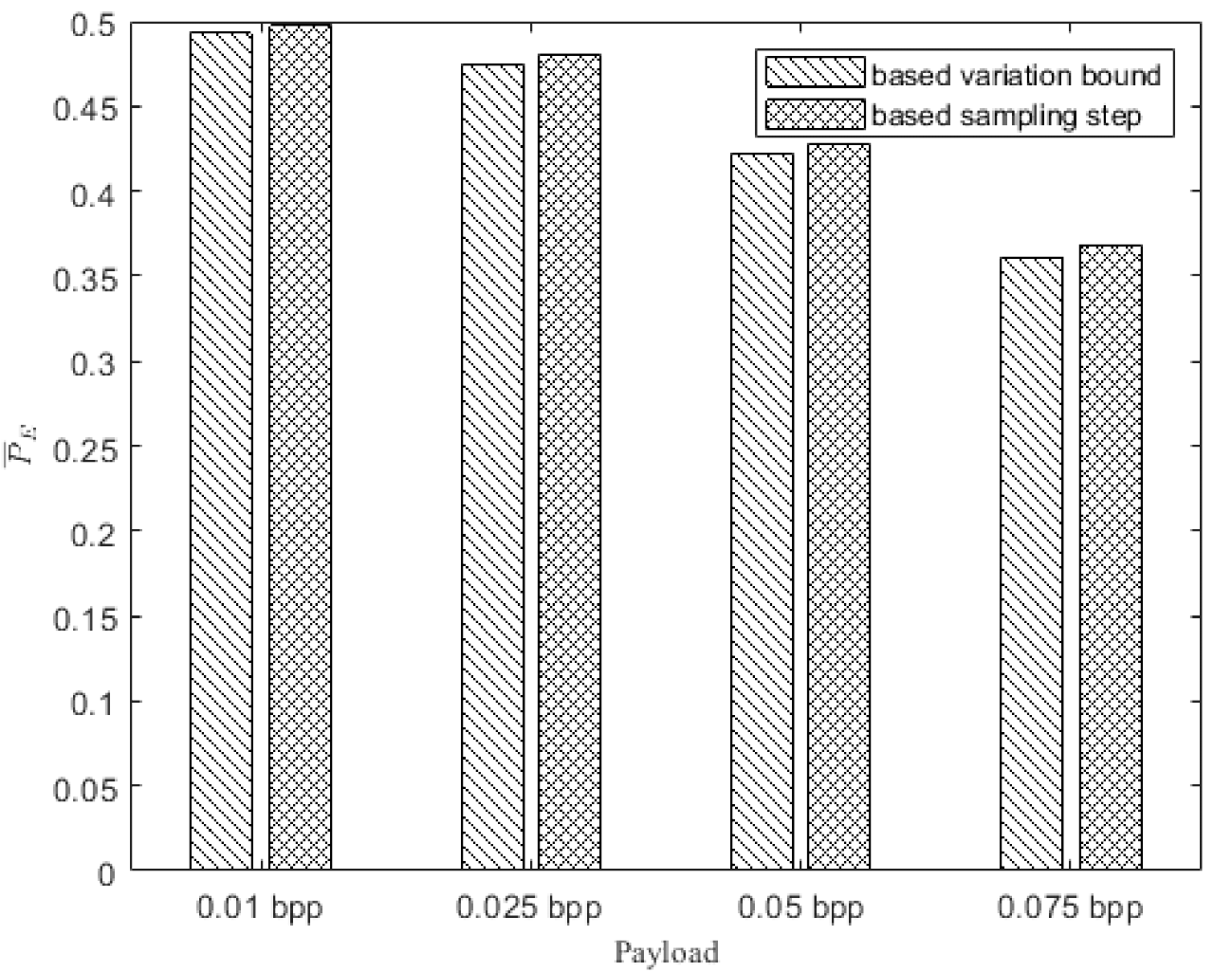,width=1\linewidth}}
\centering
\caption{The comparison of security performance in terms of $\bar{P}_{E}$ for various embedding rates within the achievable rate when implementing the S-UNIWARD-Pro with different sampling interval $s$ for an anti-aliasing bicubic interpolation scaling channel with $S\!F=0.7$.}
\label{Fig:bargrap}
\end{figure}

\subsubsection{The Discussion on the Sampling Intervals}
In our framework of robust image steganography against general scaling attacks, a unique correspondence between $\forall y_{u,v}$ in embeddable sub-image $Y\!E\!P$ and its supporting block $X\!P_{u,v}$ in cover image $X$ should be established to simplify the IP-based inverse interpolation. 
In general, sampling with interval $s$ on scaled image $Y$ should be taken to retain sufficient pixels with ${\rm dPI}=1$ in supporting block $X\!P$.  
And the number of pixels with ${\rm dPI}=1$ increases with the sampling interval $s$,  which would generally somewhat improve the security performance of the proxy stego image. 
This is because the embedding changes $\pm 1$ at $y_{u,v}$ could be allocated to more pixels $x_{i,j}$s in supporting block $X\!P_{u,v}$ during reverse interpolation, leading to less variation bound $\Delta_{u,v}$.  
We take the scaling channel of anti-aliasing bicubic interpolation with $S\!F=0.7$ as an example.
If the embedding method S-UNIWARD-Pro with design parameters in Table \ref{tab:samplestep} is adopted, the achievable embedding rate is $\sqrt{3}/9$ bpp ($\sqrt{3}/s^2$) for ternary embedding, where the sampling interval $s=3$ and $N=2$ (the number of pixels with ${\rm dPI}=1$ in $X\!P$).
Alternatively, the S-UNIWARD-Pro could also be implemented with sampling interval $s=4$ at an achievable rate of $\sqrt{3}/16$ bpp, and the pixel number with ${\rm dPI}=1$ is increased from 2 to 4.  
The security performance with the two different implementations for various embedding rates ($\leq$ the achievable rate $\sqrt{3}/16$bpp) is summarized in Fig. \ref{Fig:bargrap}. 
It is observed that increasing the sampling interval $s$ could do contribute to the improvement of security performance for the generated proxy stego images, although it is marginal on average, at the cost of less achievable embedding rate. 
Therefore, the design parameters in Table \ref{tab:samplestep}, which has a better trade-off between capacity and security, could still be taken into account in priority for practical applications of the proposed robust image steganographic scheme against scaling attacks. 

\begin{figure*}[htbp]
\centering
\subfigure[]{\label{Fig:SUNIWARD-nearest}
\includegraphics[width=0.4\linewidth]{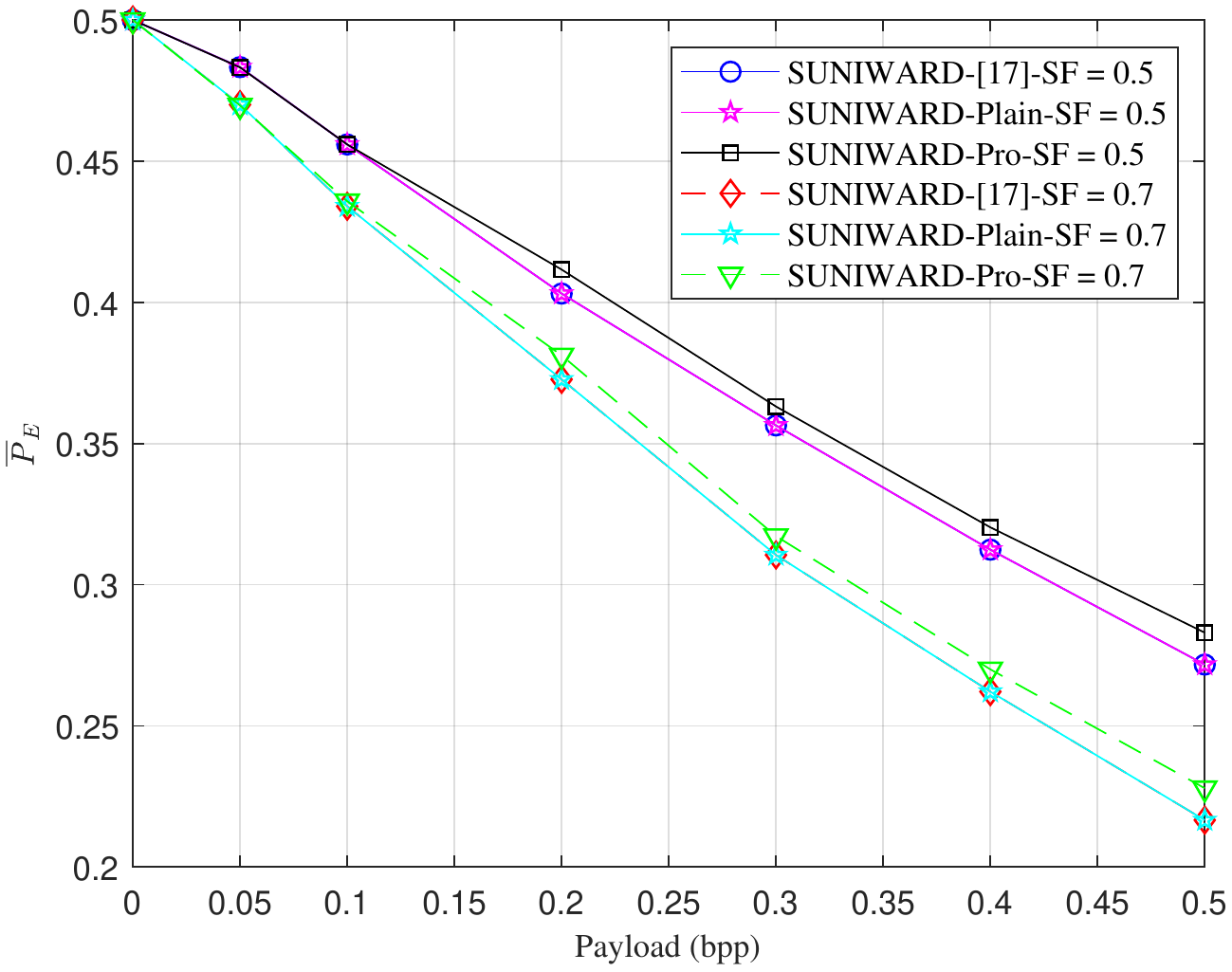}}
\subfigure[]{\label{Fig:HiLL-nearest}
\includegraphics[width=0.4\linewidth]{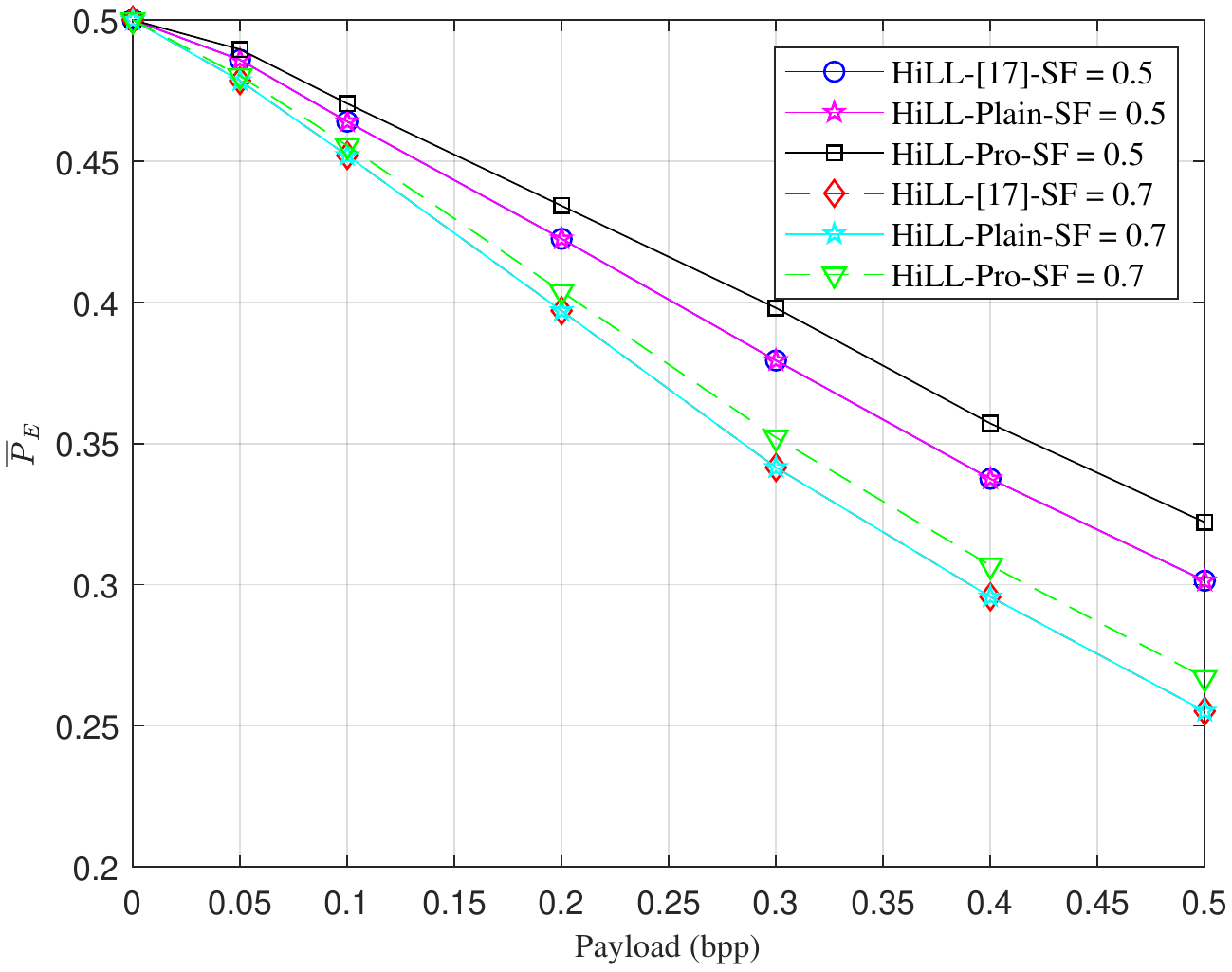}}
\\
\centering
\caption{The comparison of the proposed method with the competing method in terms of security performance between cover and proxy stego images for scaling channels with nearest neighbor interpolation at various payloads and SFs.}
\label{Fig:nearest}
\end{figure*}
\begin{figure*}[htbp]
\centering
\subfigure[]{\label{Fig:SUNIWARD-anti-alisaingBilinear}
\includegraphics[width=0.4\linewidth]{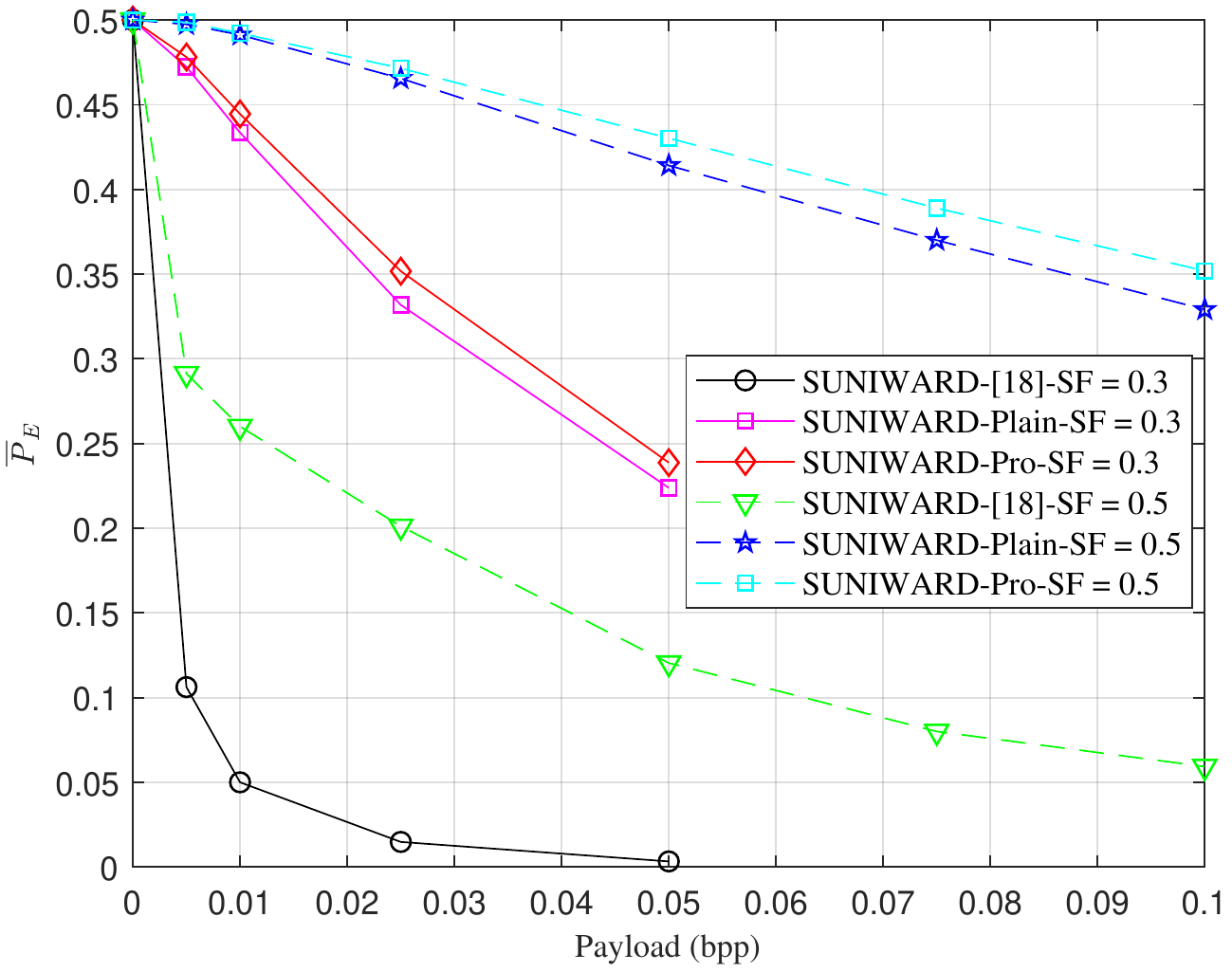}}
\subfigure[]{\label{Fig:HiLL-anti-alisaingBilinear}
\includegraphics[width=0.4\linewidth]{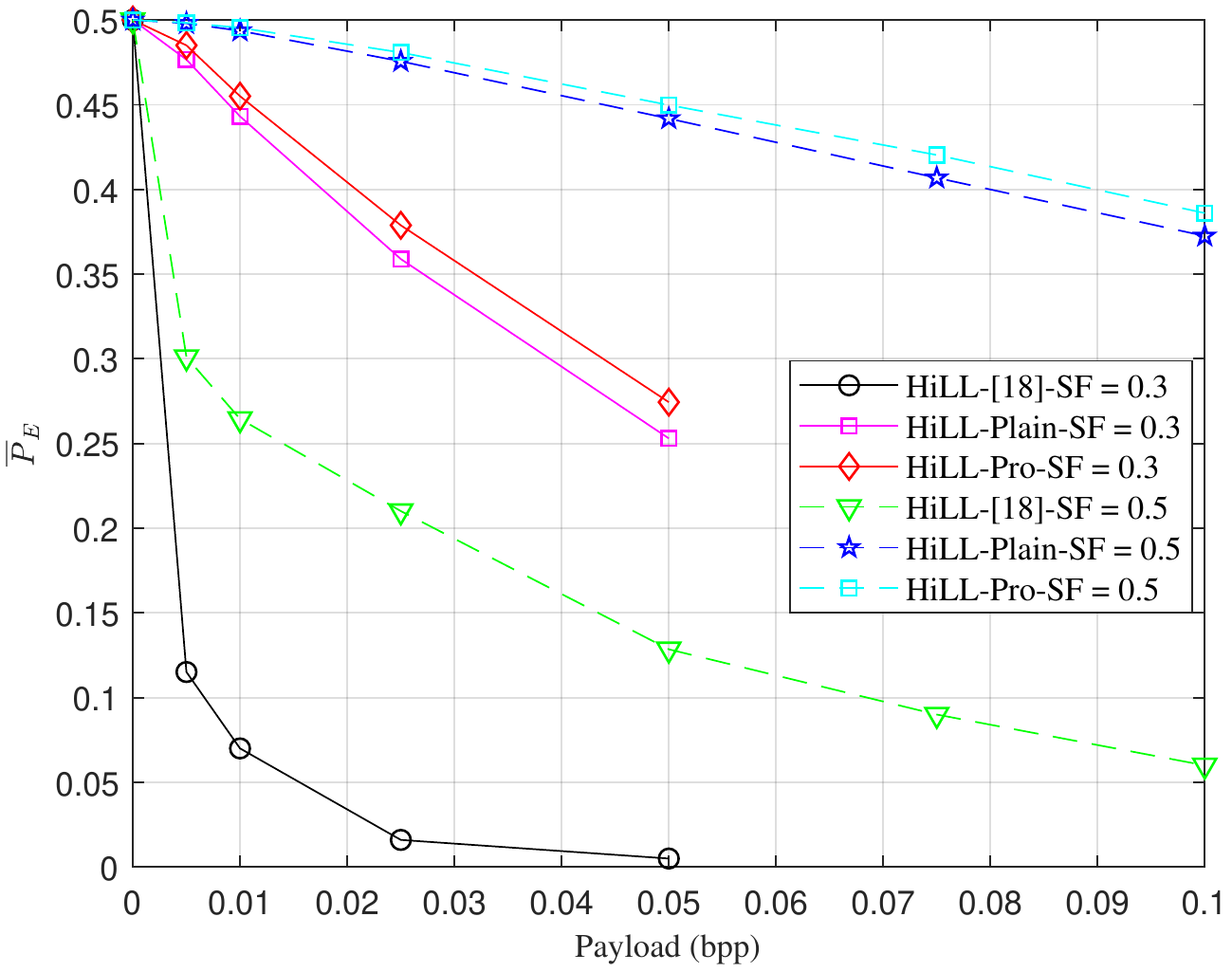}}
\\
\centering
\caption{The comparison of the proposed method with the competing method in terms of security performance between cover and proxy stego images for scaling channels with anti-aliasing bilinear interpolation at various payloads and SFs.}
\label{Fig:anti-alisaingBilinear}
\end{figure*}
\begin{figure*}[htbp]
\centering
\subfigure[]{\label{Fig:SUNIWARD-anti-alisaingBicubic}
\includegraphics[width=0.4\linewidth]{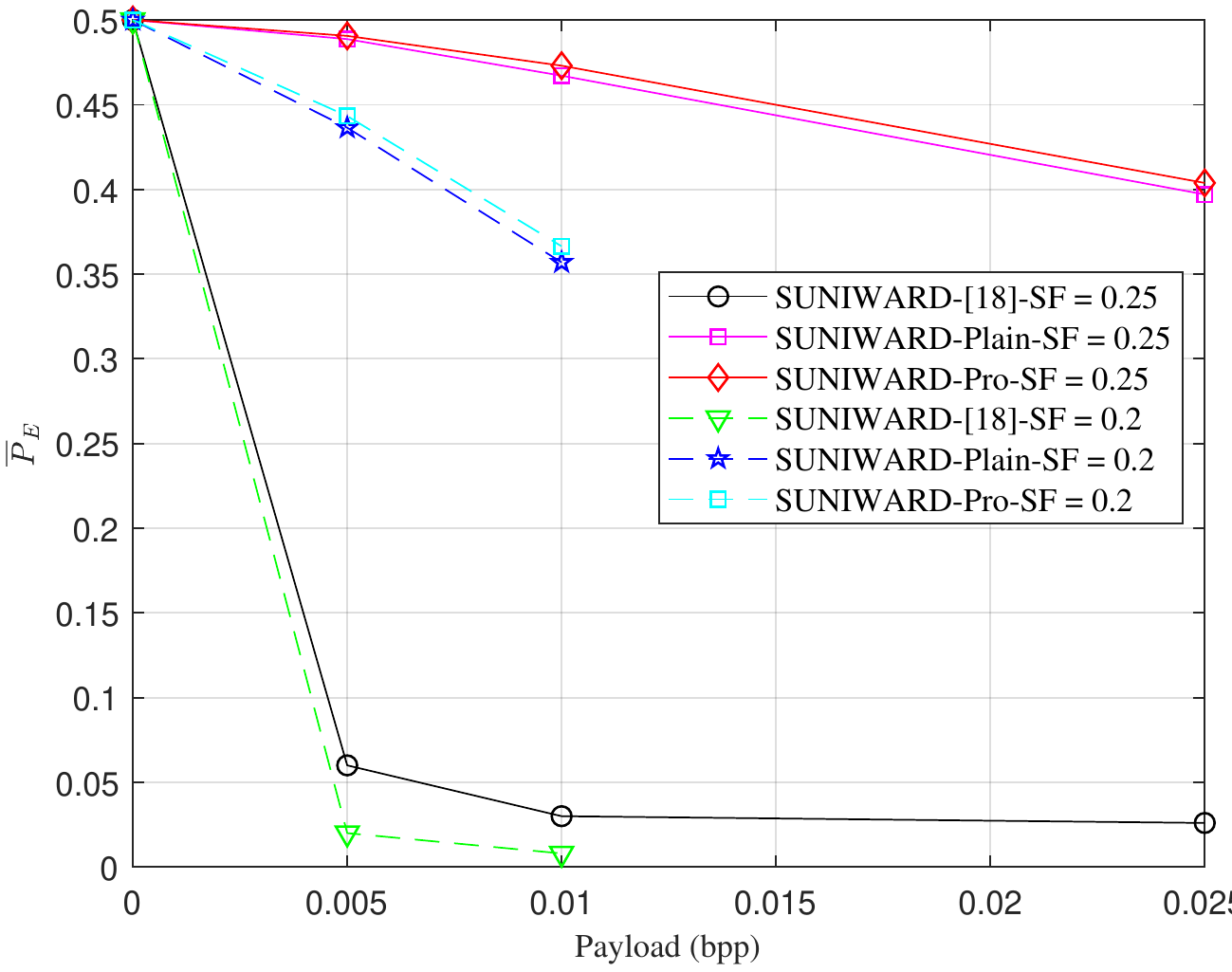}}
\subfigure[]{\label{Fig:HiLL-anti-alisaingBicubic}
\includegraphics[width=0.4\linewidth]{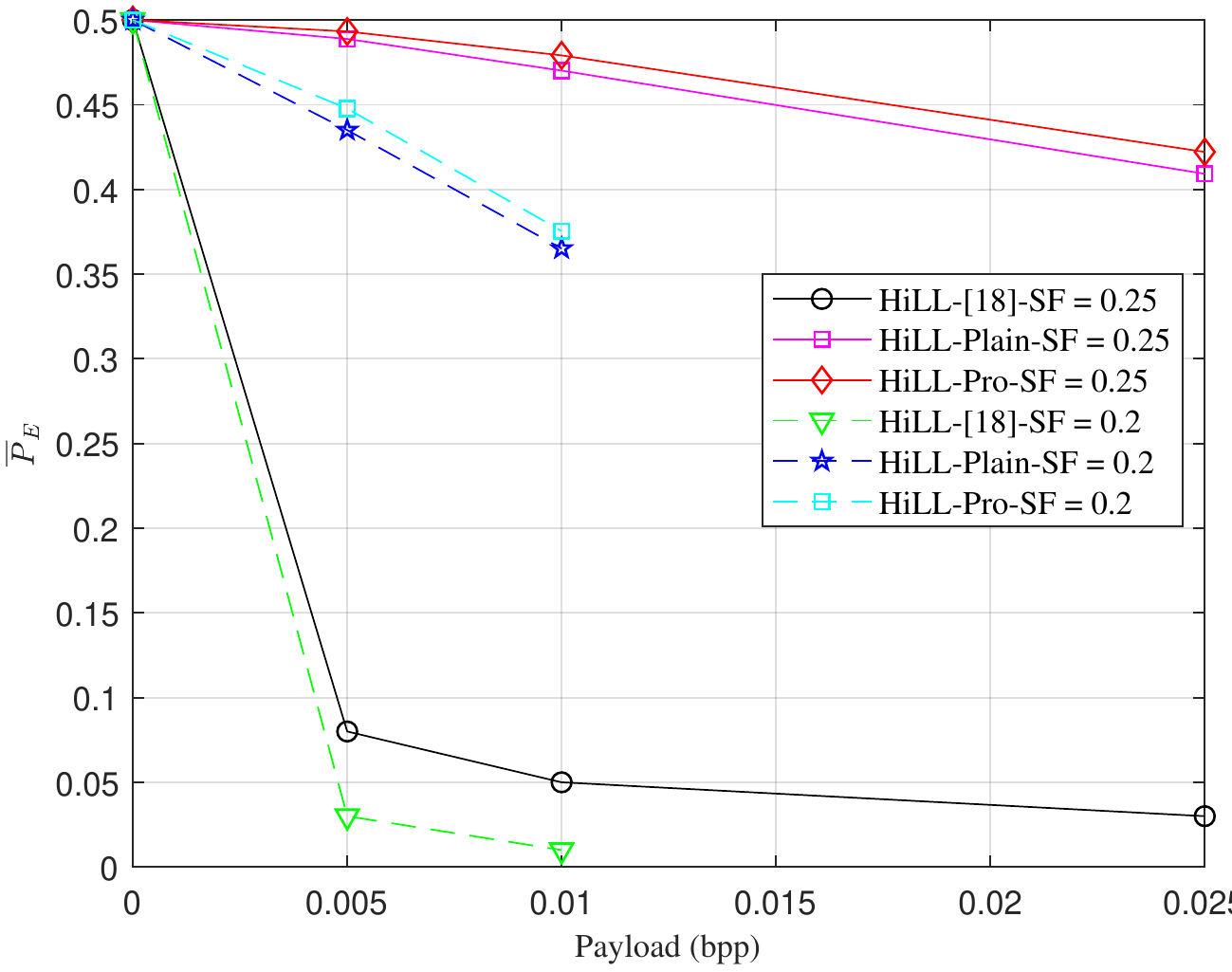}}
\\
\centering
\caption{The comparison of the proposed method with the competing method in terms of security performance between cover and proxy stego images for scaling channels with anti-aliasing bicubic interpolation at various payloads and SFs.}
\label{Fig:anti-alisaingBicubic}
\end{figure*} 
\subsection{Comparison with Prior Arts in Terms of Security performance}
We then turn to the comparison of security performance in terms of $\bar{P}_{E}$ between the proposed scheme and the prior arts for various scaling channels. 
In specific, we compare the performance of our method with the one in \cite{zhang2018image} for scaling channel with nearest neighbor interpolation, and Zhu's method \cite{zhu2021inverse} for scaling channels with both anti-aliasing bilinear and bicubic interpolations.
In the interest of fair comparison, the same embedding costs, e.g., S-UNIWARD and HiLL, and steganalyzer, e.g., SRM-34,671D, are adopted for embedding and steganalysis, respectively. 
In addition, although the proposed scheme could be generally used for the robust image steganography against the anti-aliasing interpolation scaling attacks with arbitrary SFs as shown in Table \ref{tab:security-antialiasing}, the security performance is only evaluated for the scaling channels, to which Zhu's method is applicable, i.e., anti-aliasing bilinear scaling for $S\!F$ in $(0, 0.5]$ and anti-aliasing bicubic scaling for $S\!F$ in $(0, 0.25]$. 
Unlike \cite{zhu2021inverse}, \textbf{where only the security performance for scaled stego images are evaluated}, we concentrate on the performance comparison for the proxy stego images with the identical dimension to cover images, i.e., evaluate the statistical undetectability of the proxy stego images from the cover images, which is more desirable for practical robust steganographic communications over scaling channels. For the scaling channel with nearest neighbor interpolation, Fig. \ref{Fig:nearest} shows the performance comparison of the proposed schemes (Plain- and Pro-based) with the method based on invariant pixels \cite{zhang2018image}, and the proposed method would degenerate into the one in \cite{zhang2018image} and exhibits similar security performance if a Plain-based embedding cost is adopted. 
It is also observed that our method equipped with Pro-based distortion functions consistently outperforms the one in \cite{zhang2018image} by incorporating the statistics of cover images for all the tested SFs and payloads, especially for relatively larger embedding rates, as shown in Fig. \ref{Fig:nearest}. 
As for the more challenging scaling channels with both anti-aliasing bilinear and bicubic interpolations, the proposed methods with either Plain-based or Pro-based embedding costs are indisputably superior to the ones in \cite{zhu2021inverse} by a clear margin for tested SFs and payloads. 
This is because, with the method in \cite{zhu2021inverse}, for $\forall y_{u,v}$ in scaled image $Y$, the corresponding supporting block $X\!P_{u,v}$ is usually located on the boundary of its interpolation block $P_{u,v}$ and assigned with smaller weights, thus leading to much larger variations $\Delta\!X\!P_{u,v}$ in proxy stego image $X^{\prime}$ due to the embedding change $\Delta\!y_{u,v} = \pm 1$ through inverse interpolation. 
It is worth noting that, although Zhu's method \cite{zhu2021inverse} has a maximum embedding rate of $\sqrt{3}$ bpp with ternary embedding for the scaling channels under consideration, the secure embedding payload, however, is much more desirable in practical steganographic applications.  

\section{Conclution}
The traditional image steganographic schemes are generally fragile to the lossy channels with interpolation scaling. 
In line with this, a framework for robust image steganography against general scaling attacks with either standard bilinear and bicubic interpolations and their anti-aliasing variants, is proposed in this paper. 
Following the principle of backward mapping for image interpolation scaling, the scaled image is decomposed into the embeddable sub-image ($Y\!E\!P$) and idle sub-image ($Y\!I\!P$) to establish a unique correspondence between each embeddable pixel in $Y\!E\!P$ and the supporting block centered on its interpolation block in cover image. 
And the task of robust image steganography can then be formulated as the one of constrained integer programming to obtain the proxy stego image with the identical dimension to the cover image, aiming at perfectly recover the embedded data from scaled stego image (robustness) while minimizing the $L_{1}$ norm between cover and proxy stego images (security). 
The metric - degree of pixel involvement (dPI) is adopted to identify the modifiable pixels in cover image due to the embedding changes in scaled stego image, which is incorporated to effectively solve the constrained optimization problem with branch and bound algorithm (B\&B).  
Recall that the security performance is evaluated in terms of the statistical indistinguishability of proxy stego images from cover images, the embedding costs with customized design, which are derived from the cover images rather than the scaled images themselves as done in prior arts, are also developed.  
By exploring the statistics of cover images, the proposed cost functions are shown to be capable of further boosting the security performance of the proxy stego images within the framework of minimal distortion embedding.
Extensive experimental results demonstrate that the proposed scheme could not only survive the scaling attacks with various interpolation methods at arbitrary scaling factors (SFs), but also outperforms the prior arts in terms of security by a clear margin.
\section*{ACKNOWLEDGEMENT}
The authors would like to thank Dr. Liyan Zhu at PLAIEU for providing the code of inverse interpolation in \cite{zhu2021inverse}.
\ifCLASSOPTIONcaptionsoff
  \newpage
\fi

\bibliographystyle{IEEEtran}
\bibliography{reference}

\begin{thebibliography}{10}
\providecommand{\url}[1]{#1}
\csname url@samestyle\endcsname
\providecommand{\newblock}{\relax}
\providecommand{\bibinfo}[2]{#2}
\providecommand{\BIBentrySTDinterwordspacing}{\spaceskip=0pt\relax}
\providecommand{\BIBentryALTinterwordstretchfactor}{4}
\providecommand{\BIBentryALTinterwordspacing}{\spaceskip=\fontdimen2\font plus
\BIBentryALTinterwordstretchfactor\fontdimen3\font minus
  \fontdimen4\font\relax}
\providecommand{\BIBforeignlanguage}[2]{{%
\expandafter\ifx\csname l@#1\endcsname\relax
\typeout{** WARNING: IEEEtran.bst: No hyphenation pattern has been}%
\typeout{** loaded for the language `#1'. Using the pattern for}%
\typeout{** the default language instead.}%
\else
\language=\csname l@#1\endcsname
\fi
#2}}
\providecommand{\BIBdecl}{\relax}
\BIBdecl

\bibitem{filler2011minimizing}
T.~Filler, J.~Judas, and J.~Fridrich, ``Minimizing additive distortion in
  steganography using syndrome-trellis codes,'' \emph{IEEE Transactions on
  Information Forensics and Security}, vol.~6, no.~3, pp. 920--935, 2011.

\bibitem{holub2012designing}
V.~Holub and J.~Fridrich, ``Designing steganographic distortion using
  directional filters,'' in \emph{2012 IEEE International Workshop on
  Information Forensics and Security (WIFS)}.\hskip 1em plus 0.5em minus
  0.4em\relax IEEE, 2012, pp. 234--239.

\bibitem{holub2014universal}
V.~Holub, J.~Fridrich, and T.~Denemark, ``Universal distortion function for
  steganography in an arbitrary domain,'' \emph{EURASIP Journal on Information
  Security}, vol. 2014, no.~1, p.~1, 2014.

\bibitem{li2014new}
B.~Li, M.~Wang, J.~Huang, and X.~Li, ``A new cost function for spatial image
  steganography,'' in \emph{2014 IEEE International Conference on Image
  Processing (ICIP)}.\hskip 1em plus 0.5em minus 0.4em\relax IEEE, 2014, pp.
  4206--4210.

\bibitem{Sedighi2016content}
V.~Sedighi, R.~Cogranne, and J.~Fridrich, ``Content-adaptive steganography by
  minimizing statistical detectability.'' \emph{IEEE Transactions on
  Information Forensics and Security}, vol.~11, no.~2, pp. 221 -- 234, 2016.

\bibitem{su2021image}
W.~Su, J.~Ni, X.~Hu, and J.~Fridrich, ``Image steganography with symmetric
  embedding using gaussian markov random field model.'' \emph{IEEE Transactions
  on Circuits and Systems for Video Technology}, vol.~31, no.~3, pp.
  1001--1015, 2021.

\bibitem{guo2014uniform}
L.~Guo, J.~Ni, and Y.-Q. Shi, ``Uniform embedding for efficient jpeg
  steganography.'' \emph{IEEE Transactions on Information Forensics and
  Security}, vol.~9, no.~5, pp. 814 -- 825, 2014.

\bibitem{guo2015using}
L.~Guo, J.~Ni, W.~Su, C.~Tang, and Y.-Q. Shi, ``Using statistical image model
  for {JPEG} steganography: uniform embedding revisited,'' \emph{IEEE
  Transactions on Information Forensics and Security}, vol.~10, no.~12, pp.
  2669--2680, 2015.

\bibitem{hu2018efficient}
X.~Hu, J.~Ni, and Y.-Q. Shi, ``Efficient {JPEG} steganography using domain
  transformation of embedding entropy,'' \emph{IEEE Signal Processing Letters},
  vol.~25, no.~6, pp. 773--777, 2018.

\bibitem{R2022efficient}
R.~Cogranne, Q.~Giboulot, and P.~Bas, ``Efficient steganography in jpeg images
  by minimizing performance of optimal detector.'' \emph{IEEE Transactions on
  Information Forensics and Security}, vol.~17, pp. 1328 -- 1343, 2022.

\bibitem{zhao2019improving}
Z.~Zhao, Q.~Guan, H.~Zhang, and X.~Zhao, ``Improving the robustness of adaptive
  steganographic algorithms based on transport channel matching.'' \emph{IEEE
  Transactions on Information Forensics and Security}, vol.~14, no.~7, pp. 1843
  -- 1856, 2019.

\bibitem{lu2021secure}
W.~Lu, J.~Zhang, X.~Zhao, W.~Zhang, and J.~Huang, ``Secure robust jpeg
  steganography based on autoencoder with adaptive bch encoding.'' \emph{IEEE
  Transactions on Circuits and Systems for Video Technology}, vol.~31, no.~7,
  pp. 2909 -- 2922, 2021.

\bibitem{matlabresize}
MathWorks, ``Mathworks official website - support documentation,''
  \url{https://ww2.mathworks.cn/help/matlab/ref/imresize.html?lang=en}.

\bibitem{opencvreize}
OpenCV, ``Opencv official website - geometric image transformations,''
  \url{https://docs.opencv.org/4.5.1/da/d54/group\_\_imgproc\_\_transform.html\#ga47a974309e9102f5f08231edc7e7529d}.

\bibitem{zhang2018an}
Y.~Zhang, J.~Gan, Q.~Cheng, D.~Ye, and Z.~Li, ``An image steganography
  algorithm based on quantization index modulation resisting scaling attacks
  and statistical detection.'' \emph{Computers, Materials and Continua},
  vol.~56, no.~1, pp. 151--167, 2018.

\bibitem{zhang2019zernike}
Y.~Zhang, X.~Luo, Y.~Guo, C.~Qin, and F.~Liu, ``Zernike moment-based spatial
  image steganography resisting scaling attack and statistic detection.''
  \emph{IEEE Access}, vol.~7, pp. 24\,282 -- 24\,289, 2019.

\bibitem{zhang2018image}
Y.~Zhang, X.~Luo, J.~Wang, C.~Yang, and F.~Liu, ``A robust image steganography
  method resistant to scaling and detection.'' \emph{Jounal of Internet
  Technology}, vol.~19, no.~2, pp. 607 -- 618, 2018.

\bibitem{zhu2021inverse}
L.~Zhu, X.~Luo, Y.~Zhang, C.~Yang, and F.~Liu, ``Inverse interpolation and its
  application in robust image steganography.'' \emph{IEEE Transactions on
  Circuits and Systems for Video Technology}, vol.~32, no.~6, pp. 4052 -- 4064,
  2022.

\bibitem{fridrich2012rich}
J.~{Fridrich} and J.~{Kodovsky}, ``Rich models for steganalysis of digital
  images,'' \emph{IEEE Transactions on Information Forensics and Security},
  vol.~7, no.~3, pp. 868--882, 2012.

\bibitem{Bas2011Break}
P.~Bas, T.~Filler, and T.~Pevn{\'y}, ``Break our steganographic system: the ins
  and outs of organizing {BOSS},'' in \emph{2011 Information Hiding - 13th
  International Conference(IH)}.\hskip 1em plus 0.5em minus 0.4em\relax
  Springer, 2011, pp. 59--70.

\end{thebibliography}

% \begin{IEEEbiography}[{\includegraphics[width=1in,height=1.25in,clip,keepaspectratio]{Liu}}]{Qingliang Liu}
% received the B.S. degree in Mathematics and Applied Mathematicsand in 2015 and received the M.E. degree in Statistics in 2018 from Qingdao University of Science and Technology, Qingdao, China. He is currently pursing the Ph.D. degree with the School of Data and Computer Science, Sun Yat-sen University, Guangzhou, China. His current research interests include stegonography, steganalysis and multimedia security.
% \end{IEEEbiography}

\end{document}